\documentclass[12pt]{article}
\usepackage[a4paper,textwidth=17.5cm,textheight=25cm]{geometry}
\usepackage[authoryear,semicolon]{natbib}
\usepackage{url}
\usepackage{hyperref}
\usepackage{array}
\usepackage{caption}
\usepackage[flushleft]{threeparttable}
\usepackage{multirow}
\usepackage{graphicx}
\usepackage{mathtools}
\usepackage{algpseudocode} 
\usepackage{algorithm}
\usepackage{xcolor}
\hypersetup{
    colorlinks,
    linkcolor={black!50!black},
    citecolor={blue!50!black},
    urlcolor={blue!80!black}
}
\usepackage{fixmath}
\usepackage{float}
\usepackage{amsmath,amssymb}
\usepackage{soul}
\usepackage{setspace}
\usepackage{enumerate}
\usepackage{multirow}
\usepackage{mathrsfs}
\usepackage{authblk}

\usepackage{enumerate}
\usepackage{multirow}
\usepackage{mathrsfs}
\usepackage{titlesec}
\titleformat{\section}
  {\normalfont\fontsize{12}{12}\bfseries}{\thesection}{0.5em}{}

\titleformat{\subsection}
  {\normalfont\fontsize{12}{12}\bfseries}{\thesubsection}{0.5em}{}

\titleformat{\subsubsection}
  {\normalfont\fontsize{12}{12}\bfseries}{\thesubsubsection}{0.5em}{}

\titleclass{\subsubsubsection}{straight}[\subsection]

\newcounter{subsubsubsection}[subsubsection]
\renewcommand\thesubsubsubsection{\thesubsubsection.\arabic{subsubsubsection}}

\titleformat{\subsubsubsection}
  {\normalfont\fontsize{12}{12}\bfseries}{\thesubsubsubsection}{0.5em}{}
\titlespacing*{\subsubsubsection}
{0pt}{3.25ex plus 1ex minus .2ex}{1.5ex plus .2ex}

\makeatletter
\renewcommand\paragraph{\@startsection{paragraph}{5}{\z@}%
  {3.25ex \@plus1ex \@minus.2ex}%
  {-1em}%
  {\normalfont\normalsize\bfseries}}
\renewcommand\subparagraph{\@startsection{subparagraph}{6}{\parindent}%
  {3.25ex \@plus1ex \@minus .2ex}%
  {-1em}%
  {\normalfont\normalsize\bfseries}}
\def\toclevel@subsubsubsection{4}
\def\toclevel@paragraph{5}
\def\toclevel@paragraph{6}
\def\l@subsubsubsection{\@dottedtocline{4}{7em}{4em}}
\def\l@paragraph{\@dottedtocline{5}{10em}{5em}}
\def\l@subparagraph{\@dottedtocline{6}{14em}{6em}}
\makeatother

\makeatletter
\g@addto@macro\normalsize{%
  \setlength\abovedisplayskip{-.8 pt}
  \setlength\belowdisplayskip{5pt}
  \setlength\abovedisplayshortskip{-.8 pt}
  \setlength\belowdisplayshortskip{5pt}
}
\makeatother

\begin{document}

\title{\vspace{-.5 cm} \large \textbf {Bayesian design for minimising prediction uncertainty in bivariate spatial responses with applications to air quality monitoring} \vspace{-0.4cm}} \vspace{-2.8cm}
\author[]{\normalsize Senarathne SGJ\thanks{Corresponding author: jagath.gedara@hdr.qut.edu.au}}
\author[2]{ M\"{u}ller WG}
\author[]{McGree JM}
\affil[]{\small School of Mathematical Sciences, Science and Engineering Faculty, Queensland University of Technology, Brisbane, Australia\vspace{0.3cm}}
\affil[2]{\small 
Department of Applied Statistics, Johannes Kepler University Linz, Austria}
\date{}
\maketitle \vspace{-.8cm}

\begin{center} {\textbf{ABSTRACT}} \end{center}

Model-based geostatistical design involves the selection of locations to collect data to minimise an expected loss function over a set of all possible locations. The loss function is specified to reflect the aim of data collection, which, for geostatistical studies, could be to minimise the prediction uncertainty at unobserved locations. In this paper, we propose a new approach to design such studies via a loss function derived through considering the entropy about the model predictions and  the parameters of the model. The approach also includes a multivariate extension to generalised linear spatial models, and thus can be used to design experiments with more than one response. Unfortunately, evaluating our proposed loss function is computationally expensive so we provide an approximation such that our approach can be adopted to design realistically sized geostatistical studies. This is demonstrated through a simulated study and through designing an air quality monitoring program in Queensland, Australia. The results show that our designs remain highly efficient in achieving each experimental objective individually, providing an ideal compromise between the two objectives. Accordingly, we advocate that our approach could be adopted more generally in model-based geostatistical design.


\section{Introduction}

The importance of spatial dynamics in natural processes has become a major focus of enquiry in many fields including ecology, agriculture and marine biology \citep{Bloom1999, Bruno2001,Castrignano2008, Falk2014}. Often, what can be explored and ultimately inferred about such dynamics depends on how the data were collected, and, in particular, the specific locations in space \citep{Werner2007}. In this paper, we propose an approach in Bayesian design for selecting sampling locations to efficiently learn about the spatial dynamics underpinning natural processes. In particular, we focus on quantifying and minimising uncertainty about model predictions at unobserved locations which simultaneously provides precise estimates of model parameters.

One common approach to modelling spatial dynamics is via a Gaussian process model \citep{Diggle2003}. Such a model assumes that the observations can be regarded as a realisation from a Gaussian process $\{S(\mathbold{d});\mathbold{d}\in \Re^2\}$. Typically, $S(\cdot)$ is assumed to be stationary and isotropic with mean identically equal to zero and covariance function $\text{Cov}(h)=\text{Cov}(S(\mathbold{d}),S(\mathbold{d}^{'}))$ where $h=||\mathbold{d}-\mathbold{d}^{'}||$ is the Euclidean distance between locations $\mathbold{d}$ and $\mathbold{d}^{'}$. For ease of notation, the $i$th observation $S(\mathbold{d}_i)$, where $\mathbold{d}_i$ is the location of the $i$th observation, will be abbreviated to $s_i$ throughout this article. Through such a model, one can then leverage information about the spatial variability between locations to yield predictions at unobserved locations. 

Bayesian inference provides a rigorous framework to quantify and handle uncertainty in spatial predictions, and a number of authors have considered this framework to design geostatistical studies. The work of \cite{DigglePrediction} compared the prediction performance of different classes of designs such as the `lattice plus close pairs' and `lattice plus in-fill' designs under parameter uncertainty. As such, no optimisation of the design was undertaken. This is presumably because of the large computational time involved in evaluating their loss function, and this is typical of designs for prediction. A pseudo-Bayesian design approach was proposed by \cite{Falk2014} for sampling on stream networks. Optimal designs were found under various loss functions but their work was limited to assuming some or all parameter values were known {\it a priori}. Bayesian approaches to design monitoring networks were proposed by \cite{muller2004} and \cite{Fuentes2007} such that accurate predictions could be obtained while minimising the cost of monitoring. However, both approaches do not take into account parameter uncertainty. Further, to form a dual-objective loss function, both approaches considered a linear combination of loss functions which requires specifying the relative importance of each objective via pre-defined weights.  Unfortunately, such an approach has been shown to be difficult to apply in practice \citep{Hill1968, Cook1994, McGree2008}. For a similar purpose, \cite{le2003designing} proposed an entropy-based criterion for designing networks for monitoring multivariate environmental fields. However, their approach can only be applied to network problems where the responses are approximately normally distributed, and hence has limited applicability.

Across all of the design methods proposed in the above cited papers, only a limited number of approaches have taken into account both parameter uncertainty and the uncertainty in the predicted outcomes when assessing a design. This is potentially a major limitation as both sources of uncertainty could be significant. Further, most of these approaches were limited to consider spatially dependent univariate responses or multivariate responses that are marginally Gaussian. This is potentially because of the difficulty in constructing a multivariate distribution that appropriately describes the dependencies between each response. As each response may be of a different type (i.e.\ continuous, count, binary, etc), this can lead to a rather complex model, rendering many approaches in Bayesian design computationally infeasible. Such a limitation seems rather restrictive as multiple responses are often observed in geostatistical studies e.g.\ \cite{Bohorquez2017,Musafer2017}. 

To address the limitations of previous research, we consider a generalised linear spatial modelling (GLSM) framework where the dependence between responses is described by a Copula model. Such functions are defined based on the cumulative distribution function (CDF) of the univariate responses, meaning that the joint distribution can be defined for a variety of different data types. Then, given such a model, we propose to quantify the uncertainty in spatial predictions via an entropy-based loss function.  The benefit of this is that one can then exploit the additivity property of entropy, avoiding the need to pre-specify weights on different objectives like estimation and prediction. To demonstrate our approach, we design a simulated study and also an air quality monitoring program in Queensland, Australia, and assess the performance of the resulting designs.

This paper is outlined as follows. In Section 2, the GLSM framework for modelling spatial outcomes is defined, and the Copula representation for describing multivariate spatial data is introduced. Our Bayesian design framework is described in Section 3, along with our approach to quantify and minimise uncertainty in model predictions.  In doing so, we show how such an approach also minimises uncertainty about parameter estimates, and provide an approximation to efficiently evaluate this loss function. To illustrate our methodologies, Section 4 focuses on finding designs in two examples where bivariate mixed spatial outcomes are observed. The paper concludes with a discussion of key findings and suggestions for future research.

\section{Motivating example}

Exposure to high levels of air pollution can have a variety of adverse effects on human health, including respiratory infections, heart disease, and lung cancer \citep{watson2006air,Jiang2016,zhang2016short}. This has led to an initiative by the Queensland government in Australia to monitor air quality across the state to provide information for managing and ultimately reducing air pollution. Key indicators of air quality that are measured across the state include $NO_2$ and $PM_{2.5}$ concentrations which are known to have significant adverse effects on human health \citep{ROBERTS20198,Huang2019}. However, such monitoring is hugely expensive, particularly in terms of setting up and maintaining the monitoring stations. Consequently, there is recent interest in exploring the impact of reduced sampling across the network, and how this might affect regulator's ability to predict air pollution at unsampled locations of interest. This provides motivation for the research proposed in this paper where a retrospective sampling design is sought across the current network of air quality monitoring stations in Queensland.

\subsection{Available data}

\begin{figure}[H]
\centering
\includegraphics[width=14cm]{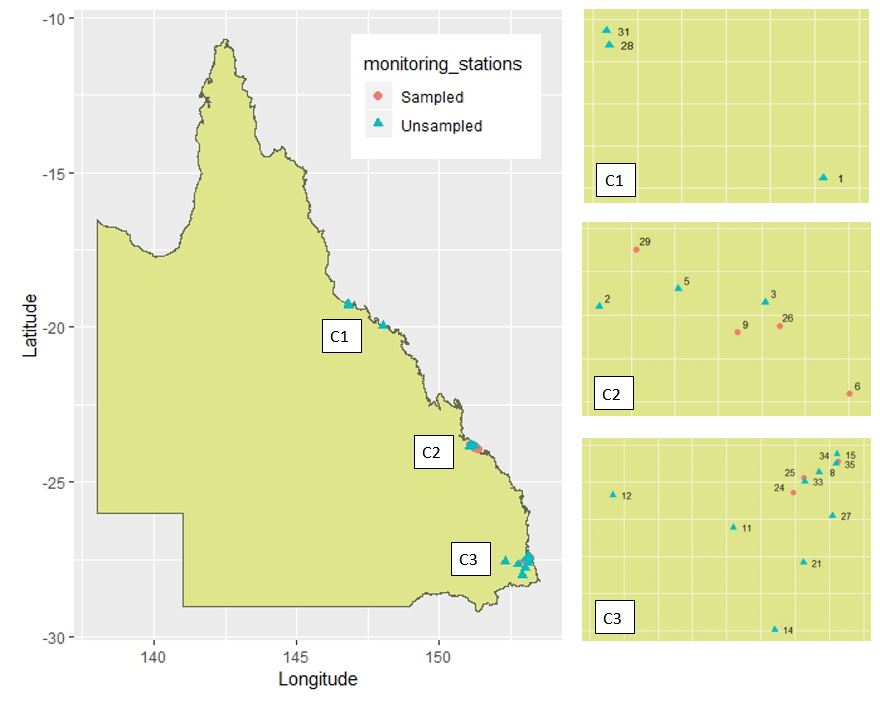} \vspace{-0.3cm}
\caption{The locations of the selected air quality monitoring stations}
\label{fig:site.loc}
\end{figure}

The Queensland air quality monitoring network consists of 22 monitoring stations; seven of which measure both $NO_2$ and $PM_{2.5}$ concentrations. Figure \ref{fig:site.loc} shows the spatial configuration of these 22 stations where we will refer to stations that measure both $NO_2$ and $PM_{2.5}$ concentrations as ``sampled" locations and stations that do not measure both of these concentrations as ``unsampled" locations.

For the sampled locations, the annual mean $NO_2$ concentrations ($Y_1$), the number of days (per annum) which exceed the daily $PM_{2.5}$ limit of $10 {\mu}gm^{-3}$ ($Y_2$), and meteorological data from year 2013 to 2016 were sourced from the Queensland government website (\sloppy \href{https://data.qld.gov.au/ dataset}{https://data.qld.gov.au/dataset}). These data were then used to develop a generalised linear spatial model to describe the observed variability in both pollutants. This model was then used to inform the selection of proposed monitoring stations within the retrospective design.
 \vspace{-0.5cm}

\section{Modelling bivariate responses in a geostatistical study}\label{modelling}

In this section, we describe our modelling framework for bivariate responses collected in a geostatistical study.  We start by first describing how a univariate response could be modelled, then extend to bivariate responses.

\subsection{Spatial model for a univariate response} 

To model a univariate response collected in a geostatistical study, we consider a GLSM which has the following form:  \vspace{.5cm}
\[
    {\mu_i}= \mathbold{X}_{i}^T\mathbold{\beta} + s_i  \hspace{.5cm} \text{and} \hspace{.5cm} E({y_i|s_i})=g^{-1}(\mu_i), \; \text{for} \; i=1,2,\ldots,n,
\]

where $\mathbold{X}_i=(1,X_{i1},\ldots,X_{ip-1})^T$ is the covariate vector associated with the location $\mathbold{d}_i$, $\mathbold{\beta} \in \Re^p$ are the regression coefficients (fixed effects), and ${s_i}$ is the value of the $i$th random effect. Here, the link function $g(.)$ defines the relationship between the linear predictor ${\mu_i}$ and the expected outcome given $s_i$. 

In a GLSM, conditional on the random effects, the dependent variable $Y$ follows a distribution from the exponential family. The joint distribution of the random effects $s_1, s_2, \ldots, s_n$ is assumed to be Multivariate Normal with mean vector zero and covariance matrix $\mathbold{\Sigma}$. When formulating this covariance matrix, it is convenient to consider a specific parametric family of covariance functions \citep{Albert1995,Diggle2003}. In this paper, the Mat\'{e}rn covariance function is used, and has the following form: \vspace{.5cm}
\[
    \text{Cov}(h;\mathbold{\gamma},\nu)= \gamma_1\rho(h,\gamma_2,\nu) \; \text{and} \; \rho(h;\gamma_2,\nu)=\frac{1}{2^{\nu-1}\Gamma(\nu)}\big(h/\gamma_2 \big)^{\nu}K_{\nu}(h/\gamma_2),
\]

where $\nu>0$, $\mathbold{\gamma}=(\gamma_1,\gamma_2)$ and $h$ is the Euclidean distance between two locations. $\Gamma(\cdot)$ is the gamma function and $K_{\nu}(\cdot)$ is the modified Bessel function of order $\nu$. The parameters $\gamma_1$ and $\gamma_2$ are the  partial sill and the spatial range, respectively. Here, we define the covariance function at the linear predictor level. Additional variability will be assumed at the data level, the form of which will depend on the data type.

To describe a Bayesian framework for the above model, let $\mathbold{y}_i=(y_{i1},\ldots,y_{in_{i}})^T$ denote the observed data at location $\mathbold{d}_i$ where $n_i$ is the number of observations collected at location $\mathbold{d}_i$, for $i=1,\ldots,n$. In spatial design, the design $\mathbold{d}=(\mathbold{d}_1,\ldots,\mathbold{d}_n)^T \in \mathscr{D}$ represents the locations where the outcomes $\mathbold{y}=(\mathbold{y}_1,\ldots,\mathbold{y}_n)^T$ are measured, and $\mathscr{D}$ represents the design space for the study. Let $p(\mathbold{\theta})$ denote the prior distribution about the parameters $\mathbold{\theta}$. Then, within a Bayesian framework, all inferences are based on the posterior distribution defined as follows:\vspace{.5cm}
\[
    p(\mathbold{\theta|y,d}) \propto {\int\limits_{\mathbold{s \in S}} \prod\limits_{i=1}^n \prod\limits_{j=1}^{n_i} p(y_{ij}|\mathbold{d}_i,s_i,\mathbold{\beta})p(\mathbold{s|\gamma}, \nu) p(\mathbold{\theta}) \text{d}\mathbold{s}},
\]

where $\mathbold{s}=(s_1,\ldots,s_n)^T$ is vector of random effects, and the parameter vector $\mathbold{\theta}$ includes both model parameters $\mathbold{\beta}$ and covariance parameters $\mathbold{\gamma}$ and $\nu$. $p(y_{ij}|\mathbold{d}_i,s_i,\mathbold{\beta})$ is the conditional likelihood of observing $y_{ij}$ at location $\mathbold{d}_i$ given the model parameters $\mathbold{\beta}$ and random effects $s_i$ for $i=1,\ldots,n$ and $j=1,\ldots,n_i$.

To extend the above model to handle bivariate responses, we consider Copula models to describe the dependence between responses \citep{nelsen2006introduction,Bardossy2006,KAZIANKA2011310,GRALER2011206} (see next section). Adopting such an approach provides a framework to flexibly model dependence between bivariate responses. 

\subsection{Spatial model for bivariate responses} \label{SP_cop_model}

Here, we define a modelling approach for bivariate spatial responses by combining the univariate spatial models (defined above) via a suitable Copula model. For this, assume that two response variables will be observed with one being continuous $Y_1$ and the other being discrete $Y_2$; thus marginally they can be modelled via two GLSMs as follows: \vspace{.5cm}
\begin{equation*}
    {\mu_{1i}}= \mathbold{X}_{i}^T\mathbold{\beta}_1 + s_{1i}, \hspace{.2cm} E({y_{1i}|s_{1i}})=g_1^{-1}(\mu_{1i}),  \hspace{.2cm} \text{and}
    \label{eq:GLMM1} 
\end{equation*} 
\begin{equation*}
    {\mu_{2i}}= \mathbold{X}_{i}^T\mathbold{\beta}_2 + s_{2i}, \hspace{.2cm} E({y_{2i}|s_{2i}})=g_2^{-1}(\mu_{2i}) \; \text{for} \; i=1,2,\ldots,n.
    \label{eq:GLMM2}
\end{equation*}

\noindent We assume that $Y_1$ and $Y_2$ have marginal probability distributions $f_{Y_1|s_{1i}}$ and $f_{Y_2|s_{2i}}$ given the random effects $s_{1i}$ and $s_{2i}$, respectively. Further, denote the marginal cumulative distribution function of $Y_1|s_{1i}$ and  $Y_2|s_{2i}$ as $F_{Y_1|s_{1i}}$ and $F_{Y_2|s_{2i}}$, respectively. Then, the Copula representation of the joint distribution $G_{Y_1|s_{1i},Y_2|s_{2i}}$ is given by, \vspace{.5cm}
\begin{equation*} 
\begin{split}
G_{Y_1|s_{1i},Y_2|s_{2i}}(y_{1ij},y_{2ij}) & =C\big(F_{Y_1|s_{1i}}(y_{1ij}), F_{Y_2|s_{2i}}(y_{2ij});\alpha\big) \\
& =C(u_{1ij},u_{2ij};\alpha) \; \text{for} \; i=1,2,\ldots,n \; \text{and} \; \text{for} \; j=1,2,\ldots,n_i,  
\end{split}
\end{equation*} 

where $C$ and $\alpha$ denote the Copula function and the Copula parameter, respectively. The Copula parameter $\alpha$ defines the strength of dependence within the bivariate distribution \citep{nelsen2006introduction}.

From this, the Copula representation of the joint distribution of $Y_1|s_{1i}$ and $Y_2|s_{2i}$ is given by, \vspace{.5cm}
\begin{equation}
f_{Y_1|s_{1i},Y_2|s_{2i}}(y_{1ij},y_{2ij})=f_{Y_1|s_{1i}}(y_{1ij})(c_{1ij}-c_{1ij}^*),
\label{Eq:Ch4_Copula}
\end{equation}

where $c_{1ij}= \frac{\partial C(u_{1ij},u_{2ij};\alpha)}{\partial{u_{1ij}}}$ , $c_{1ij}^*= \frac{\partial C(u_{1ij},u_{2ij}^-;\alpha)}{\partial{u_{1ij}}}$ and ${u_{2ij}^-}$ is the left hand limit of $u_{2ij}$, see \cite{joe2015} and \cite{tao2013dose} for further details.

Then, using Equation (\ref{Eq:Ch4_Copula}), the conditional likelihood of a bivariate mixed outcome $(\mathbold{y}=({y_{1ij},y_{2ij}})$ for $i=1,2,\ldots,n \; \text{and} \; j=1,2,\ldots,n_i)$ can be expressed as follows: \vspace{.5cm}
\[
    p(\mathbold{y|d},\mathbold{\theta},\mathbold{s}_1,\mathbold{s}_2)= \prod\limits_{i=1}^n \prod\limits_{j=1}^{n_i} \Big [\big(f_{Y_1|s_{1i}}(y_{1ij})\big)\big(c_{1ij}-c_{1ij}^*\big)\Big],
\]

where $\mathbold{\theta}$ includes all the model parameters ($\mathbold{\beta}_1,\mathbold{\beta}_2$), the covariance parameters ($\mathbold{\gamma}_1=( \gamma_{11}, \gamma_{21}),$ $\mathbold{\gamma}_2=( \gamma_{21}, \gamma_{22}), \nu_1, \nu_2$), and the Copula parameter $\alpha$.

The bivariate Copula function describes the dependence structure between two random variables. A large number of bivariate Copulas and their dependence properties have been discussed in the literature \citep{durante2010copula,genest1986joy,nelsen2006introduction}. Among them, the bivariate Archimedean Copula models such as Clayton, Gumbel and Frank Copulas have been widely used due to their flexibility in describing different tail behaviours and different dependence structures. 

The Clayton Copula is considered in this paper to model the joint distribution of the responses with the reader referred to \cite{clayton1978model} and \cite{Cook1981} for further details. Let $Y_1$ and $Y_2$ be two responses with CDFs $u_1=F_{Y_1}(y_1)$ and $u_2=F_{Y_2}(y_2)$, respectively, then the joint CDF of  $Y_1$ and $Y_2$ can be expressed using the Clayton Copula as follows: \vspace{.5cm}
\[
C(u_1,u_2;\alpha) = ({u_1}^{-\alpha}+{u_2}^{-\alpha}-1)^{-1/\alpha};    \hspace{0.5cm}    \alpha > 0,
\]

where $\alpha$ is the Copula parameter.

The choice of this Copula model was motivated by the tail dependence structure of the data observed in the motivating example and desirable properties of the Archimedean Copula family to which this Copula belongs. For example, as provided by \cite{genest1986joy}, there is a closed form relationship between the bivariate Archimedean Copula parameter and Kendall's tau ($\tau$). Therefore, once the Copula parameter is estimated, it is straightforward to define the dependence between the two responses within the intuitive scale of $-1$ to $+1$. For example, the relationship between the Clayton Copula parameter $\alpha$ and Kendall's tau can be defined as follows: \vspace{.5cm}
\[
\tau=4E[C(u_1,u_2;\alpha)]-1=\frac{\alpha}{\alpha+2}.
\]

\section{Bayesian design framework for geostatistical studies} \label{Sec3}

Our aim of minimising the uncertainty in spatial predictions can be quantified within a Bayesian design framework by a loss function which we will denote as $\lambda(\mathbold{d,\theta,y})$.  Such a loss function compares a summary of the posterior distribution for $\mathbold{\theta}$ (based on observing $\mathbold{y}$ from $\mathbold{d}$) with its true value.  However, as this function depends on $\mathbold{\theta}$ and $\mathbold{y}$, which are unknown {\it a priori}, it cannot be used to select designs.  Accordingly, the expectation of the loss function is used to locate designs, and this expectation can be defined as follows: \vspace{.5cm}
\begin{equation}
    L(\mathbold{d})=E_{\mathbold{y,\theta}}[\lambda(\mathbold{d,\theta,y})]= \int\limits_\mathbold{Y}\int\limits_\mathbold{\Theta} \lambda(\mathbold{d,\theta,y})p(\mathbold{y|\theta,d})p(\mathbold{\theta})\text{d}\mathbold{\theta}\text{d}\mathbold{y},
    \label{Eq:Loss_f1}
\end{equation}

where $p(\mathbold{y}|\mathbold{\theta,d}) = \int\limits_{\mathbold{s} \in \mathbold{S}} p(\mathbold{y}|\mathbold{\beta,s,d})p(\mathbold{s}|\mathbold{\gamma},\nu)d\mathbold{s}$. 

When the loss function does not depend on the model parameters $\mathbold{\theta}$, Equation (\ref{Eq:Loss_f1}) can alternatively be expressed as: \vspace{.5cm}
\begin{equation}
    L(\mathbold{d})=E_{\mathbold{y}}[\lambda(\mathbold{d,y})]= \int\limits_\mathbold{Y} \lambda(\mathbold{d,y})p(\mathbold{y|d})\text{d}\mathbold{y}.
    \label{Eq:Loss_f2}
\end{equation}

A Bayesian design is then found by minimising the expected loss over the space of all possible locations.  However, in general, the expected loss function does not have a closed-form solution, and hence, needs to be approximated. Monte Carlo (MC) integration is the most commonly used approach for this approximation. This is achieved by simulating a large number of prior predictive data sets, evaluating the loss function for each data set, and then taking the average as the approximation to the expected loss. Formally: \vspace{.5cm}
\begin{equation}
    \hat{L}(\mathbold{d})={\frac{1}{K}} \sum\limits_{k=1}^{K} \lambda(\mathbold{d},\mathbold{\theta}_k,\mathbold{y}_k),
    \label{Eq:Monte_loss}
\end{equation}

where $\mathbold{\theta}_k$ and $\mathbold{y}_k$ are generated from the distributions $p(\mathbold{\theta})$ and $p(\mathbold{y}|\mathbold{d},\mathbold{\theta}_k)$, respectively.

The current approach to Bayesian design for spatial prediction is based on a loss function proposed in \cite{DigglePrediction}. This loss function quantifies the spatially averaged prediction variance of the unobserved random field $S(\cdot)$ over the predicted region $\mathscr{A}$ as follows: \vspace{.5cm}
\begin{equation}
    \lambda_{pred}(\mathbold{d,y})=\int\limits_{\mathbold{\xi}\in\mathscr{A}}{\text{Var}\{S(\mathbold{\xi})|\mathbold{y,d}\}}\text{d}\mathbold{\xi}.
\label{Eq:Diggle_ute}    
\end{equation}

When the prediction region $\mathscr{A}$ consists a discrete set of locations $\mathbold{\xi}_{1},\mathbold{\xi}_2,\ldots,\mathbold{\xi}_T$, the above loss function can be expressed as follows: \vspace{.5cm}
\[
    \lambda_{pred}(\mathbold{d,y})=\frac{1}{T}\sum\limits_{t=1}^T{\text{Var}\{S(\mathbold{\xi}_t)|\mathbold{y,d}\}}.
\]

As discussed in  \cite{DigglePrediction}, this loss function provides some provision to also address the objective of parameter estimation. In the next section, we describe an alternative approach for addressing these dual objectives. 

\subsection{An entropy-based loss function for spatial prediction}

Here, we derive a loss function to quantify uncertainty in a spatial process by considering the entropy about the predictions, following the principles outlined in the seminal book by \citet{lezidek2006}. For this, we note that, for a given model, there are two sources of uncertainty about the predictions: (1) Uncertainty in the parameter values; and (2) Uncertainty in the outcome $\mathbold{Z}$ conditional on the parameter values. Thus, to derive this loss function, we start by considering the entropy in these two random variables {\it a priori}. Initially we consider a univariate outcome $\mathbold{Z}$ where the entropy about this random variable and the parameter values is: \vspace{.5cm}
\begin{equation*}
\begin{split}
    H(\mathbold{Z,\theta|\xi}) = - \int\limits_{\mathbold{z}\in\mathcal{Z}}\int\limits_{\mathbold{\Theta}} p(\mathbold{z,\theta|\xi})\log p(\mathbold{z,\theta|\xi}) \text{d}\mathbold{\theta}\text{d}\mathbold{z}, \; \text{and} \\  H(\mathbold{Z,\theta|\xi}) = - \sum\limits_{\mathbold{z}\in\mathcal{Z}}\int\limits_{\mathbold{\Theta}} p(\mathbold{z,\theta|\xi})\log p(\mathbold{z,\theta|\xi}) \text{d}\mathbold{\theta},
\end{split}    
\end{equation*}

for cases where $\mathbold{Z}$ is a continuous and discrete outcome, respectively. 

Following this, we define the loss function in terms of the change in entropy about $\mathbold{Z}$ and the parameters upon observing data $\mathbold{y}$ at design $\mathbold{d}$ as follows: \vspace{.5cm}
\begin{equation}
    \lambda_D(\mathbold{d,y})= H(\mathbold{Z,\theta|y,d,\xi})- H(\mathbold{Z,\theta|\xi}),
\label{Eq:Dual_ute1}
\end{equation}

where $H(\mathbold{Z,\theta|y,d,\xi})$ is the entropy about $\mathbold{Z}$ and the parameters upon observing data $\mathbold{y}$ at design $\mathbold{d}$.

Using the chain rule of entropy, it is straightforward to show that the joint entropy of $\mathbold{Z}$ and the parameters is equal to the conditional entropy of $\mathbold{Z}$ given the parameters plus the entropy of the parameters. That is: \vspace{.5cm}
\begin{equation}
\begin{split}
    H(\mathbold{Z,\theta|y,d,\xi})= H(\mathbold{Z|\theta,y,d,\xi})+H(\mathbold{\theta|y,d}), \; \text{and} \;
    H(\mathbold{Z,\theta|\xi})= H(\mathbold{Z|\theta,\xi})+H(\mathbold{\theta}),
\end{split} 
\label{Eq:cond_entrop}
\end{equation}

where, when $\mathbold{Z}$ is a discrete random variable, $H(\mathbold{Z|\theta,\xi})= -\sum\limits_{\mathbold{z}\in\mathcal{Z}}\int\limits_{\mathbold{\Theta}} p(\mathbold{z,\theta|\xi})\log \frac{p(\mathbold{z,\theta|\xi})}{p(\mathbold{\theta})} \text{d}\mathbold{\theta}$ and $H(\mathbold{\theta})=-\int\limits_{\mathbold{\Theta}} p(\mathbold{\theta})\log {p(\mathbold{\theta})} \text{d}\mathbold{\theta}$.

Then, by substituting the above expressions into Equation (\ref{Eq:Dual_ute1}), the loss function $\lambda_D(\mathbold{d,y})$ can be expressed as follows: \vspace{.5cm}
\begin{equation}
\begin{split}
    \lambda_D(\mathbold{d,y}) & = H(\mathbold{Z,\theta|y,d,\xi})- H(\mathbold{Z,\theta|\xi}) \\
    & = \big \{H(\mathbold{Z|\theta,y,d,\xi})+H(\mathbold{\theta|y,d}) \big\}- \big\{H(\mathbold{Z|\theta,\xi})+H(\mathbold{\theta}) \big\}\\
    & = \big \{H(\mathbold{Z|\theta,y,d,\xi})- H(\mathbold{Z|\theta,\xi}) \big\}+ \big\{ H(\mathbold{\theta|y,d}) - H(\mathbold{\theta}) \big\}\\
    & = \lambda_P(\mathbold{d,y})+\lambda_E(\mathbold{d,y}).
\end{split}
\label{Eq:Dual_ute2}
\end{equation}

As shown in Equation (\ref{Eq:Dual_ute2}), the loss function $ \lambda_D(\mathbold{d,y})$ can be expressed as a sum of two loss functions in which the first ($ \lambda_P(\mathbold{d,y})$) quantifies the change in entropy about $\mathbold{Z}$ given the parameters $\mathbold{\theta}$ while the second ($\lambda_E(\mathbold{d,y})$) quantifies the change in entropy about the parameter values. Thus, the loss function $\lambda_D(\mathbold{d,y})$ is termed as a dual-purpose loss function for parameter estimation and prediction. The derivation of Equation (\ref{Eq:Dual_ute2}) is essentially the same as the fundamental identity given in \citet{lezidek2006}.

For parameter estimation, the loss function $\lambda_E(\mathbold{d,y})$ is equivalent to the Kullback-Leibler divergence (KLD) between the prior and the posterior distributions of parameters \citep{kullback1951information}, and can be expressed as follows: \vspace{.5cm}
\begin{equation}
   \lambda_E(\mathbold{d,y})= - \int\limits_\mathbold{\Theta}{p(\mathbold{\theta|y,d})}\log\Bigg(\frac{p(\mathbold{\theta|y,d})}{p(\mathbold{\theta})}\Bigg) \text{d}\mathbold{\theta}.
    \label{Eq:Ch4_Ute_est1}
\end{equation}

When comparing designs via the loss function $\lambda_P(\mathbold{d,y})$, the term $H(\mathbold{Z|\theta,\xi})$ is independent of the design $\mathbold{d}$. As such, for simplicity, we consider  $H(\mathbold{Z|\theta,y,d,\xi})$ as our loss function for prediction given the parameters as follows: \vspace{.5cm}
\begin{equation}
 \lambda_P(\mathbold{d,y}) = H(\mathbold{Z|\theta,y,d,\xi})  = - \int\limits_{\mathbold{\Theta}}p(\mathbold{\theta|y,d})\sum\limits_{\mathbold{z}\in\mathcal{Z}}p(\mathbold{z|\theta,y,d,\xi})\log p(\mathbold{z|\theta,y,d,\xi}) \text{d}\mathbold{\theta}.
\label{Eq:Ud_Pr1}
\end{equation}

However, when assessing the performance of designs under this expected loss, the term $H(\mathbold{Z|\theta,\xi})$ is subtracted so that the resulting values are interpretable.

When a bivariate mixed outcome (i.e.\ one outcome is continuous and the other is discrete) is observed ($\mathbold{z}=(\mathbold{z_1,z_2})$), the above loss function can be expressed as follows: \vspace{.5cm}
\begin{equation}
\lambda_P(\mathbold{d,y}) = - \int\limits_{\mathbold{\Theta}}p(\mathbold{\theta|y,d}) \sum\limits_{\mathbold{z_2}\in\mathcal{Z}_2} \int\limits_{\mathbold{z_1}\in \mathcal{Z}_1}{p(\mathbold{z_1,z_2|\theta,y,d,\xi})}
 {\log p(\mathbold{z_1,z_2|\theta,y,d,\xi})}\text{d}\mathbold{z_1}\text{d}\mathbold{\theta},
\label{Eq:Ud_Pr2}
\end{equation}

where $\mathbold{y}=(\mathbold{y_1,y_2}) $ contains data collected for the two responses at the locations $\mathbold{\xi}$ based on design $\mathbold{d}$.

The loss function $\lambda_E(\mathbold{d,y})$ for the bivariate mixed outcome is defined as above, where $\mathbold{\theta}$ contains all model parameters for both responses.

\section{Efficient approximations to loss functions} \label{Sec:Des_algorithm}

Evaluating the MC approximation to the expected loss in Equation (\ref{Eq:Ud_Pr2}) requires approximating or sampling from a large number of posterior distributions.  Unfortunately, this renders algorithms like Markov chain Monte Carlo computationally infeasible to use in locating designs. Accordingly, in this section, we describe computationally efficient methods for approximating the above loss functions.

\subsection{Approximating the posterior distribution}

To efficiently evaluate the MC approximation to the expected loss function in Equation (\ref{Eq:Ud_Pr2}), fast methods for approximate inference are required.  For this purpose, we consider the Laplace approximation to the posterior distribution, see \cite{Long201324,overstall2017}. For this, we assume, approximately, that:  \vspace{.5cm}
\[
    (\mathbold{\theta|y,d})\sim MVN(\mathbold{\theta}^*,\mathbold{A}(\mathbold{\theta}^*)^{-1}),
\]

where $\mathbold{\theta}^*=\underset{\mathbold{\theta}}{\operatorname{arg\,max}}\{\log p(\mathbold{y|d,\theta})+\log p(\mathbold{\theta})\}$ and $\mathbold{A}(\mathbold{\theta}^*)$ is the negative Hessian matrix: \vspace{.5cm}
\[           
       \mathbold{A}(\mathbold{\theta}^*) = \frac{-\partial^2\big\{\log p( \mathbold{y|d,\theta})+\log p(\mathbold{\theta})\ \big\}}{\partial\mathbold{\theta}\partial\mathbold{\theta}^{'}}\Big|_{\mathbold{\theta}=\mathbold{\theta}^*}. 
\] 

In our spatial model, the linear predictor contains random effects, and these need to be integrated out in order to evaluate the likelihood. That is: \vspace{.5cm}
\begin{equation*}
\begin{split}
    p(\mathbold{y|d,\theta})=\idotsint\limits_{\mathbold{S_1}}\idotsint\limits_{\mathbold{S_2}}\prod\limits_{i=1}^{n}\prod\limits_{j=1}^{n_i}\Big [\big(f_{Y_1|s_{1i}}(y_{1ij})\big)\big(c_{1ij}-c_{1ij}^*\big)\Big] p(\mathbold{s}_1|\mathbold{\gamma}_1, \nu_1) \times \\
     p(\mathbold{s}_2|\mathbold{\gamma}_2, \nu_2) \text{d}{s_{11}}\ldots\text{d}{s_{1n}}\text{d}{s_{21}}\ldots\text{d}{s_{2n}}.
\end{split}    
\end{equation*}

Unfortunately, given the form of our model, there will typically be no closed-form solution to the above integral, and therefore the likelihood.  To handle this, we again employ MC integration to approximate this integral by simulating random effects as follows $s_{1ib}\sim p(s_{1i}|\mathbold{\gamma}_1, \nu_1),\; s_{2ib}\sim p(s_{2i}|\mathbold{\gamma}_2, \nu_2) \;\text{for}\; i=1,2,\ldots,n \;\text{and}\; b=1,2,\ldots,B$, and we note that a variety of approximations could be used, see for example \cite{Gomez-Rubio2019}. Using MC integration, the likelihood can be approximated as follows: \vspace{.5cm}
\begin{equation}
\begin{split}
    p(\mathbold{y|d,\theta}) & \approx \frac{1}{B}\sum\limits_{b=1}^B \prod\limits_{i=1}^n\prod\limits_{j=1}^{n_i} p(y_{ij}|d_{i},\mathbold{\theta},s_{1ib},s_{2ib}). 
\end{split}    
    \label{Eq:like_approx}
\end{equation}

It is this approximation to the likelihood that will be used to form a Laplace approximation to the posterior distribution (as shown above).

\subsection{Approximating the loss function}

When both the prior and posterior distributions follow Multivariate Normal distributions with mean vectors $(\mathbold{\theta}_0,\mathbold{\theta}^*)$ and covariance matrices $(\mathbold{\Omega}_0,\mathbold{\Omega})$, respectively, the loss function $\lambda_E(\mathbold{d,y})$ can be evaluated as follows: \vspace{.5cm}
\begin{equation}
    \tilde{\lambda}_E(\mathbold{d,y})= -\frac{1}{2}\Bigg( tr\big(\mathbold{\Omega}_0^{-1}\mathbold{\Omega}\big) +(\mathbold{\theta}^*-\mathbold{\theta}_0)^T\mathbold{\Omega}_0^{-1}(\mathbold{\theta}^*-\mathbold{\theta}_0)-k+\log\Big(\frac{\text{det}\mathbold{\Omega}_0}{\text{det}\mathbold{\Omega}} \Big)\Bigg),
    \label{Eq:Ch4_Ute_est2}
\end{equation}

where $k$ is the dimension of the two Multivariate Normal distributions.

In terms of approximating ${\lambda}_P(\mathbold{d,y})$, this proves to be a little more complicated as summations need to be taken across simulated values of $\mathbold{\theta}$, $\mathbold{y}$ and $\mathbold{z}$.  That is: \vspace{.5cm}
\begin{equation}
\hat{\lambda}_P(\mathbold{d,y}) = - \frac{1}{K}\sum\limits_{k=1}^K \frac{1}{R}\sum\limits_{r=1}^R\Bigg[\log \Big ( \frac{1}{B}\sum\limits_{b=1}^B {p(\mathbold{z}_r|\mathbold{s}_b,\mathbold{\theta}_k,\mathbold{y},\mathbold{d},\mathbold{\xi})}\Big )\Bigg],
\label{Eq:Monte_dualUte}
\end{equation}

where $\mathbold{\theta}_k\sim p(\mathbold{\theta|y,d})$, $\mathbold{s}_b\sim p(\mathbold{s}|\mathbold{\theta}_k\mathbold{,y,d})$  and $\mathbold{z}_r\sim p(\mathbold{z}|\mathbold{s}_b,\mathbold{\theta}_k \mathbold{,y,d,\xi})$, and the posterior distribution is approximated via a Laplace approximation.

In using this approach, in order to precisely estimate this loss function, a sufficiently large number of samples should be taken from the posterior distributions of the parameters, random effects, and the posterior predictive distribution (i.e.\ all $B$, $K$ and $R$ should be sufficiently large). Thus, evaluating this loss function is hugely expensive computationally.  Indeed, if each integral requires $N$ MC samples, the computation time is of order $N^3$ ($\mathcal{O}(N^3)$).  Unfortunately, this renders such an approach computationally infeasible for use in locating designs (even when adopting a Laplace approximation to the posterior distribution). To overcome this, we consider forming a Multivariate Normal approximation to the joint distribution of simulated data as the entropy for such a distribution has a closed form. We propose forming this approximation via moment matching. That is, in the two examples considered in this paper, a bivariate mixed outcome is observed where a Normal response ($Y_1$) and a Poisson response ($Y_2$) are considered. Given the mean value for the Poisson response is reasonably large, the joint distribution of $Y_1$ and $Y_2$ given $\mathbold{d}$ and $\mathbold{\xi}$ can be approximated by a bivariate Normal distribution. Given this, the entropy of bivariate Normal response was considered to approximate the loss function given in Equation (\ref{Eq:Ud_Pr2}), and therefore the first term of the total loss function given in Equation (\ref{Eq:Dual_ute2}) was approximated as follows: \vspace{.5cm}
\begin{equation}
    \tilde{\lambda}_P(\mathbold{d,y}) = \frac{1}{2} \sum\limits_{\mathbold{\xi}_t\in\mathbold{\xi}}\log \det(2{\pi}e \mathbold{\hat{\Sigma}}_{\mathbold{\xi}_t})=T(1+\log(2\pi))+\sum\limits_{\mathbold{\xi_t}\in\mathbold{\xi}}\log \det(\mathbold{\hat{\Sigma}}_{\mathbold{\xi}_t}),
\label{Eq:Approx_UteD}
\end{equation}

where $\mathbold{\hat{\Sigma}}_{\mathbold{\xi}_t}$ is the approximated variance-covariance matrix of the bivariate Normal distribution at location $\mathbold{\xi}_t$, for $t=1,2,\ldots,T$. Here, $\mathbold{\hat{\Sigma}}_{\mathbold{\xi}_t}$ was approximated by taking a sufficiently large sample from the posterior predictive distribution $p(\mathbold{z|s,\theta,y,d,}\mathbold{\xi}_t)$.  Thus, this approximation has complexity of order $N$ i.e.\ $\mathcal{O}(N)$, making it substantially more efficient to compute than the approximation in Equation (\ref{Eq:Monte_dualUte}).  We explore the accuracy and computational benefits of this approximation in Section 5.

Given the above two approximations, the dual-purpose loss function can be approximated as follows: \vspace{.5cm}
\begin{equation}
    \tilde{\lambda}_D(\mathbold{d,y})=\tilde{\lambda}_P(\mathbold{d,y})+\tilde{\lambda}_E(\mathbold{d,y}).
    \label{Eq:Dual_approx}
\end{equation}

All the loss functions discussed above depend on either a set of discrete choices for $\mathbold{\xi}$ or the space of $\mathbold{\xi}$.  Throughout this paper, we consider a set of discrete choices, and thus we average the utility across these when approximating the expected utility. If we were to consider a space for $\mathbold{\xi}$, then the utility would need to be integrated over this space. In cases where there is no analytic solution to this integration, then one could approximate the solution via averaging over a random selection of locations.

\subsection{Design algorithm}

With the use of the Laplace approximation for approximating the posterior distribution  and the above approximations to the expected loss, we propose Algorithm 1 to derive optimal designs for spatially dependent bivariate outcomes described by the Copula-based GLSM. 

\begin{algorithm}[H]
 \caption{Approximating the expected loss function for the location of geostatistical designs}
 	\begin{algorithmic}[1]
 	    \State Initialise the prior information $p(\mathbold{\theta})$ and the prediction set $\mathscr{A}$. 
        \For{$k=1$ to $K$}
            \State Draw $\mathbold{\theta}_k=(\mathbold{\beta}_{1k},\mathbold{\beta}_{2k},\mathbold{\gamma}_{1k},\mathbold{\gamma}_{2k},\nu_{1k},\nu_{2k},\alpha_k) \sim p(\mathbold{\theta})$ 
            \State Draw $\mathbold{s}_k=(\mathbold{s}_{1k},\mathbold{s}_{2k}) \sim p(\mathbold{s}|\mathbold{\gamma}_{1k},\mathbold{\gamma}_{2k},\nu_{1k},\nu_{2k})$ 
            \State Simulate data $\mathbold{y}_{k}$ at the design $\mathbold{d}$ from the assumed Copula model $\mathbold{y}_{k} \sim p(\mathbold{y}|\mathbold{s}_{k},\mathbold{\theta}_{k},\mathbold{d})$
            \State Form Laplace approximation via $\mathbold{\theta}^*_k=\underset{\mathbold{\theta}}{\operatorname{arg\,max}}\{\log p(\mathbold{y}_k|\mathbold{\theta,d})+\log p(\mathbold{\theta})\}$ such that \newline
            \hspace*{1em} the posterior distribution can be approximated by $MVN(\mathbold{\theta}^*_k,\mathbold{A}^{-1}_k)$, where \newline
            \hspace*{1em} $\mathbold{A}_k$ is the Hessian matrix defined as: 
                \[            
                    \mathbold{A}_k = \frac{-\partial^2\big\{\log p(\mathbold{y}_k|\mathbold{\theta,d})+\log p(\mathbold{\theta})\ \big\}}{\partial\mathbold{\theta}\partial\mathbold{\theta}^{'}}\Big|_{\mathbold{\theta}=\mathbold{\theta}^*}. 
                \]
             \State Evaluate ${\lambda}_E(\mathbold{d},\mathbold{y}_k)$ via Equation (\ref{Eq:Ch4_Ute_est2}) where $\mathbold{\Omega} = \mathbold{A}^{-1}_k$
             \State Simulate data $\mathbold{z}_b$ at $\mathbold{\xi}$ via $\mathbold{z}_b\sim p(\mathbold{z}|\mathbold{s}_b,\mathbold{\theta}_b,\mathbold{\xi})$, where $\mathbold{\theta}_b = (\mathbold{\beta}_{1b},\mathbold{\beta}_{2b},\mathbold{\gamma}_{1b},\mathbold{\gamma}_{2b},\nu_{1b},\nu_{2b},\alpha_b) \sim$ \hspace*{1.2em}  $p(\mathbold{\theta}|\mathbold{y}_k,\mathbold{d})$ and $\mathbold{s}_b \sim p(\mathbold{s}|\mathbold{\gamma}_{1b},\mathbold{\gamma}_{2b},\nu_{1b}, \mathbold{v}_{2b}),$ for  $ = 1,\ldots,B$
         \State Approximate  ${\lambda}_P(\mathbold{d},\mathbold{y}_k)$ via Equation (\ref{Eq:Approx_UteD})
		  \State Approximate  ${\lambda}_D(\mathbold{d},\mathbold{y}_k)$ via Equation  (\ref{Eq:Dual_approx})
        \EndFor
        \State Approximate the expected loss $\hat{L}(\mathbold{d})=\frac{1}{K}\sum\limits_{k=1}^K {\lambda}_D(\mathbold{d},\mathbold{y}_k)$
        \State Find optimal design  $\mathbold{d}^*=\underset{\mathbold{d \in\mathscr{D}}}{\operatorname{arg\,min}}\, \hat{L}(\mathbold{d})$
 	\end{algorithmic}
 \end{algorithm} 
 
Implementing this algorithm requires defining a model for analysis, the prior information about the parameters and the prediction region (line 1). Once defined, at each iteration, a parameter $\mathbold{\theta}_k=(\mathbold{\beta}_{1k},\mathbold{\beta}_{2k},\mathbold{\gamma}_{1k},\mathbold{\gamma}_{2k},\nu_{1k},\nu_{2k},\alpha_k)$ is drawn from the prior distribution (line 3). Then,  $(\mathbold{\gamma}_{1k},\mathbold{\gamma}_{2k},\nu_{1k},\nu_{2k})$, a random effect $\mathbold{s}_k$ is simulated (line 4). Next, data $\mathbold{y}_k$ are simulated from the assumed Copula model (line 5). The posterior distribution of the parameters is then approximated via the Laplace approximation (line 6). This requires finding the posterior mode and evaluating the Hessian matrix at this mode. Here, we found the mode via a gradient-based optimisation algorithm, and the Hessian matrix was evaluated numerically. When the dual-purpose loss function is considered for design selection, $\lambda_E(\mathbold{d},\mathbold{y}_k)$ and $\lambda_P(\mathbold{d},\mathbold{y}_k)$ in lines 7 and 9 of Algorithm 1 can be approximated via Equations (\ref{Eq:Ch4_Ute_est2}) and (\ref{Eq:Approx_UteD}), respectively. The approximation of $\lambda_P(\mathbold{d},\mathbold{y}_k)$ requires generating data from the posterior predictive distribution as shown in line 8 of Algorithm 1. Following this, the dual-purpose loss function can be approximated via Equation  (\ref{Eq:Dual_approx}) (line 10). Then, once $K$ iterations have been completed, the MC approximation to the expected loss is evaluated in line 12. To find the optimal set of locations, a suitable optimisation algorithm is implemented to minimise the approximate expected loss function (line 13).\vspace{-0.2 cm}
 
\section{Case studies} \label{Sim.study} \vspace{-0.2 cm}

A simulated example and the motivating example described in Section 2 were considered to demonstrate the performance of Algorithm 1 and the dual-purpose loss function derived in Section \ref{Sec3} to design geostatistical studies. In both examples, bivariate mixed spatial outcomes are observed. As such, we first model the univariate spatial outcomes via a GLSM, and then combine these via the Clayton Copula model as described in Section \ref{SP_cop_model}.

In Example 1, we simulate data within a unit square with a Mat\'{e}rn covariance function under parameter uncertainty. As designs can vary depending upon the strength of spatial dependence, three different settings (weak, moderate and strong) were considered. In Example 2, we apply the proposed approach to find a retrospective design for the Queensland air quality monitoring network.

In both of these examples, we consider the case where a practitioner needs to select the most appropriate set of locations to collect data from within a region or finite set of locations. To explore this, we evaluated the effectiveness of selecting dual-purpose designs compared to designs that only address a single aim. For this, we first obtained prediction only and estimation only designs using the loss functions given in Equation (\ref{Eq:Diggle_ute}) and Equation (\ref{Eq:Ch4_Ute_est1}), respectively. Then, the dual-purpose designs were evaluated against these single-purpose designs to assess performance. As Example 1 considers a continuous design space, the approximate coordinate exchange (ACE) algorithm \citep{overstall2017approximate} was used to find optimal designs. Here, default settings as detailed in \cite{overstall2018ACEBAYES} were used. As Example 2 was undertaken across a restricted design space with only a fixed number of locations being available, the standard coordinate exchange algorithm \citep{Meyer1995} was used to obtain optimal designs. All simulations were run using R 3.5.2, and R code to reproduce the results in this paper is available via the following GitHub  repository, \href{https://github.com/SenarathneSGJ/Model-based_geostatistical_design}{https://github.com/SenarathneSGJ/Model-based{\textunderscore}geostatistical{\textunderscore}design}.

\subsection{Example 1: Exploring dual-purpose design for spatial processes}

In this example, consider collecting bivariate outcomes across a spatial process. Here, $Y_1$ was assumed to follow a Normal distribution given the random effects $s_1$ as follows:  \vspace{.5cm}
\begin{equation}
Y_{1}|s_1\sim N(\mu_1,\sigma_1^2) \hspace{0.5 cm} \text{and} \hspace{0.5 cm} \mu_1= \beta_{10}+\beta_{11}X_1+\beta_{12}X_2+s_1,
\label{Eq:Ex1R1}
\end{equation}

where $(\beta_{10},\beta_{11},\beta_{12})$ are the model parameters, $X_1$ and $X_2$ are the two covariates in the model and $\sigma_1^2$ is the residual variance of $Y_1$. The random effects $s_1$ were assumed to follow a Multivariate Normal distribution with mean $\textbf{0}$ and covariance matrix $\mathbold{\Sigma}_1$, where the elements of $\mathbold{\Sigma}_1$ were obtained via a Mat\'{e}rn covariance function with parameters $(\mathbold{\gamma}_1, \nu_1)$ as described in Section \ref{modelling}.

$Y_2$ was assumed to follow a Poisson distribution given the random effects $s_2$ as follows: \vspace{.5cm}
\begin{equation}
Y_{2}|s_2\sim Pois(\exp(\mu_2)) \hspace{0.5 cm} \text{and} \hspace{0.5 cm} \mu_2= \beta_{20}+\beta_{21}X_1+\beta_{22}X_2+s_2,
\end{equation}

where $(\beta_{20},\beta_{21},\beta_{22})$ are the model parameters, and $X_1$, $X_2$ are two covariates of the model. Again, the random effects $s_2$ were assumed to follow a Multivariate Normal distribution with mean $\textbf{0}$ and covariance matrix $\mathbold{\Sigma}_2$. The elements of the covariance matrix $\mathbold{\Sigma}_2$ were obtained from a Mat\'{e}rn covariance function with parameters $(\mathbold{\gamma}_{2},\nu_2)$.

Next, the model for the bivariate responses $Y_1$ and $Y_2$ was obtained using the Clayton Copula function as detailed in Section 3. It was assumed that accurate predictions were required at 25 locations ($\mathbold{\xi}$) defined as follows: \vspace{.5cm}
\[
\xi_{vw}= (0.25v,0.25w) \; \text{for} \; v,w= 0,1,\ldots,4.
\]

\begin{figure}[H]
	\centering
\includegraphics[width=11 cm]{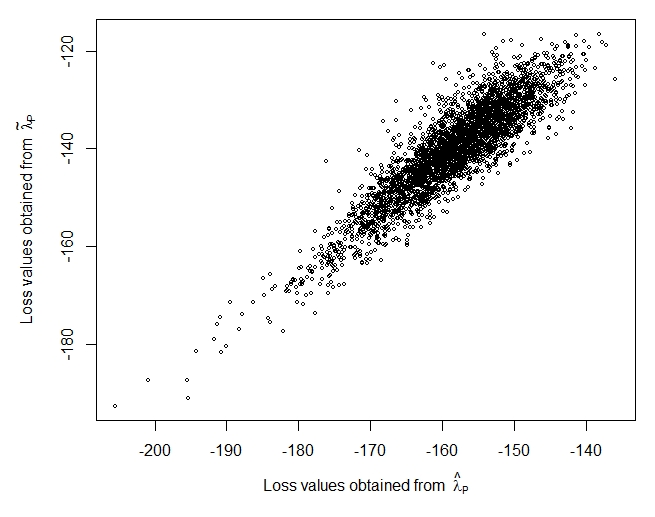} \vspace{-0.3cm}
	\caption{The relationship between the loss values obtained from $\hat{\lambda}_P(\mathbold{d,y})$ and $\tilde{\lambda}_P(\mathbold{d,y})$ for 100 randomly generated designs under moderate spatial dependence from Example 1}
	\label{fig:App_Ute_Ex1}
\end{figure}

To investigate the performance of our design algorithm and the dual-purpose loss function for obtaining designs under various spatial conditions, three design scenarios were considered. These design scenarios differ in terms of the strength of the spatial covariance, constructed via three different values for the prior mean of the spatial range parameter as given in Table \ref{Tab:design_scenarios}. 
\begin{table}[H]
\begin{center}
\caption{Strength of the covariance in three scenarios}
\label{Tab:design_scenarios}
\begin{tabular}{clc}
\hline
\multicolumn{1}{l}{Scenario} & Strength of dependence & \multicolumn{1}{l}{\begin{tabular}[c]{@{}l@{}}Prior mean of \\ range parameter ($a$)\end{tabular}} \\
 \hline
 1 & Weak & $0.2$  \\
 2 & Moderate & $0.5$ \\
 3 & Strong & $0.8$ \\
 \hline
\end{tabular}
\end{center}
\end{table} \vspace{-.5cm}
The prior distributions of the remaining parameters were same across all the scenarios. For all the model and covariance parameters, Normal priors were considered as defined in Table \ref{Tab:Priors}. Further, we assumed that there is a positive dependence between the two responses, and therefore, a Normal prior was considered for the parameter $\text{logit}(\tau)$, which we also estimated within our framework. 
\begin{table}[H]
\begin{center}
\caption{Prior distributions for all parameters} 
\label{Tab:Priors}
\begin{threeparttable}
\begin{tabular}{ l l l l} \hline 
 Parameter & Prior distribution & Parameter & Prior distribution\\[0.5ex]  \hline 
 $\beta_{10}$ & $N\big(5,4\big)$ & $\beta_{20}$ & $N\big(3.8,0.125\big)$ \\ [0.4ex] 
 $\beta_{11}$ & $N\big(-2.8,4\big)$ & $\beta_{21}$ & $N\big(-0.5,0.125\big)$ \\ [0.4ex]  
 $\beta_{12}$ & $N\big(8,4\big)$ & $\beta_{22}$ & $N\big(-0.7,0.125\big)$ \\ [0.4ex]
 $\log(\sigma_1)$ & $N\big(\log(1.2),0.25\big)$ & $\log(\gamma_{21}/\gamma_{22})$ & $N\big(\log(0.6/a),0.125\big)$ \\ [0.4ex]
 $\log(\gamma_{11}/\gamma_{12})$ & $N\big(\log(0.7/a),0.25\big)$ & $\log(\gamma_{22})$ & $N\big(\log(a),0.125\big)$\\ [0.4ex]
  $\log(\gamma_{12})$ & $N\big(\log(a),0.25\big)$ & $\log(\nu_{2})$ & $N\big(\log(0.25),0.25\big)$\\ [0.4ex] 
 $\log(\nu_{1})$ & $N\big(\log(1.5),0.25\big)$ & $\text{logit}(\tau)$ & $N\big(0.85,0.25\big)$\\ [0.4ex] 
 \hline 
\end{tabular} \vspace{0.0 cm}
    \begin{tablenotes}
        \item \small*The parameter $a\in\{0.2,0.5,0.8\}$.
    \end{tablenotes}
\end{threeparttable}
\end{center}
\end{table} 
\textbf{Results:} We first explore the accuracy of the proposed approximation to the loss function ${\lambda}_P(\mathbold{d,y})$ in Equation (\ref{Eq:Approx_UteD}). To do this, we evaluated $\tilde{\lambda}_P(\mathbold{d,y})$ and  $\hat{\lambda}_P(\mathbold{d,y})$ for 100 randomly generated designs each with 10 design points under moderate spatial dependence. The results of this are shown in Figure \ref{fig:App_Ute_Ex1} where a clear but noisy relationship between the two loss functions can be observed. Despite the noise, there is generally agreement between the two approximations, particularly in terms of designs that minimise each loss function. Similar results were also observed under weak and strong spatial dependence. As such, we propose that our approximation can be used to select designs for this example.

\begin{figure}[H]
	\centering
\includegraphics[width=11.5 cm]{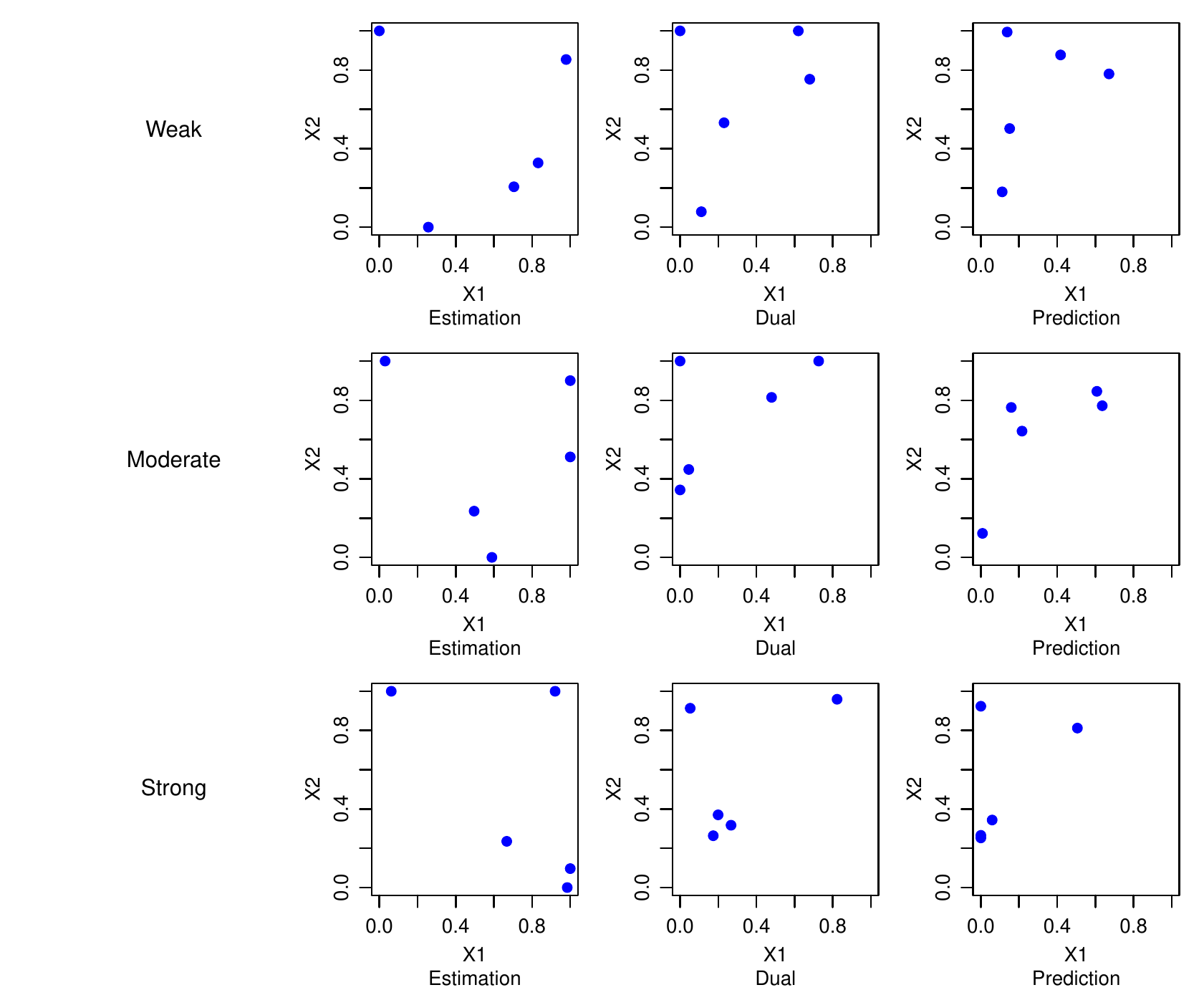} \vspace{-.2cm}
	\caption{The optimal designs selected from each loss function (n=5)}
	\label{fig:opt_d5}
\end{figure}

\begin{figure}[H]
	\centering
\includegraphics[width=11.5 cm]{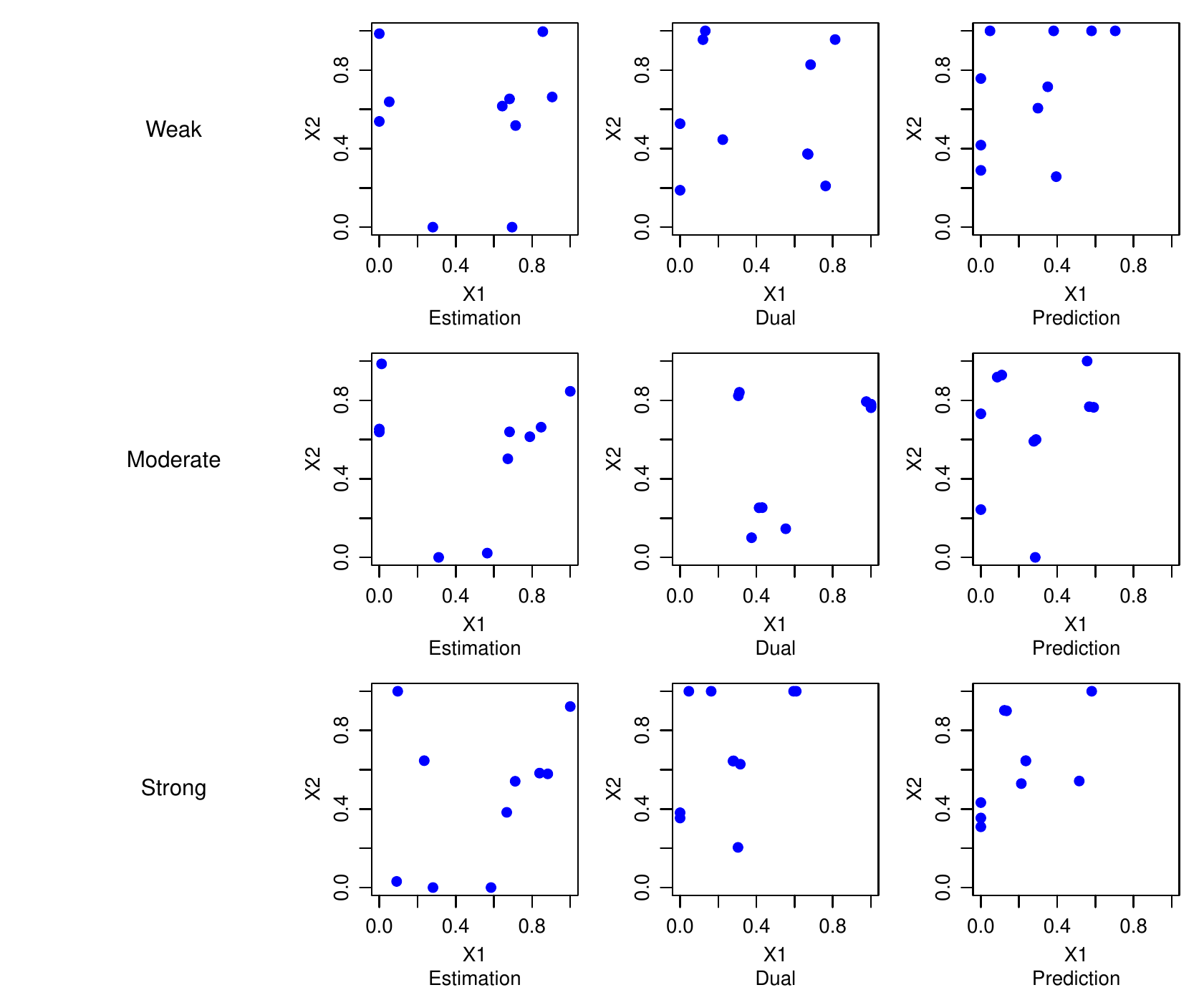} \vspace{-.2cm}
	\caption{The optimal designs selected from each loss function (n=10)}
	\label{fig:opt_d10}
\end{figure}

Figures \ref{fig:opt_d5} and \ref{fig:opt_d10} compare the optimal designs selected under the three scenarios with 5 and 10 locations, respectively. As can be seen, when there is moderate or strong spatial covariance, the designs selected under the three loss functions have different spatial configurations. A detailed comparison of these designs with those that might be suggested in the literature for estimation and prediction is given in Appendix B. Furthermore, as shown in Figure \ref{fig:opt_d10}, this example yielded some replicate points which may provide additional information on variance parameters. However, in real spatial settings, it may not be realistic to have such replicate points so we note that it would be straightforward to extend the adopted optimisation algorithm to avoid these.

\begin{figure}[H]
	\centering
\includegraphics[width=13.5cm]{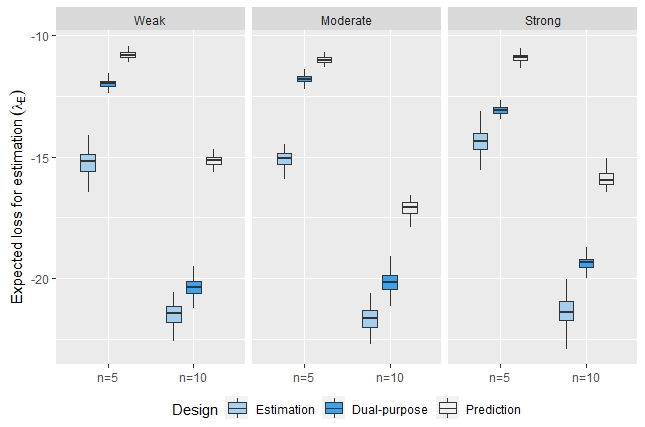} \vspace{-.3cm}
	\caption{The boxplot of the distribution of the expected values for the loss function that focuses on parameter estimation for different designs over 100 simulations in Example 1}
	\label{fig:Est_E1}
\end{figure}

\begin{figure}[H] 
	\centering
\includegraphics[width=13.5cm]{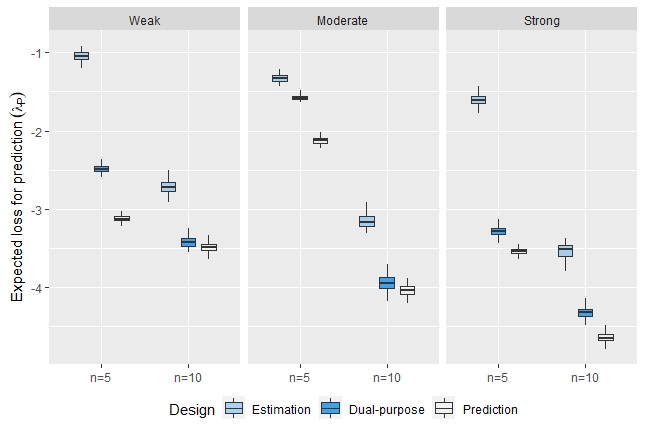} \vspace{-.4cm}
	\caption{The boxplot of the distribution of the expected values for the loss function that focuses on prediction for different designs over 100 simulations in Example 1}
	\label{fig:log_det_E1}
\end{figure}

Once the optimal designs had been found for each scenario, they were evaluated for the goals of parameter estimation and spatial prediction. For this evaluation, 100 independent estimates of the approximate loss were used. In each simulation, for a given optimal design, the expected values of the loss functions $\lambda_E(\mathbold{d,y})$ and $\lambda_P(\mathbold{d,y})$ were estimated using the MC integration as shown in Equation (\ref{Eq:Monte_loss}). This yielded a distribution of the expected loss values for each optimal design (see Figure \ref{fig:Est_E1} and Figure \ref{fig:log_det_E1}). Further, to quantitatively compare designs, a design efficiency was calculated. For this purpose, the average of the 100 expected loss values was used as follows: \vspace{.5cm}
\[
    \text{Eff}(\mathbold{d},\mathbold{d}^*_{\phi})=\frac{ \sum_{k=1}^{100} E_{\mathbold{y}_{k}}[\lambda_{\phi}(\mathbold{d},\mathbold{y}_k)]}{\sum_{k=1}^{100}E_{\mathbold{y}_{k}}[\lambda_{\phi}(\mathbold{d}^*_{\phi},\mathbold{y}_k)]}, \; \text{for} \; \phi\in\{E,P\}, 
\]

where $\mathbold{d}_E^{*}$ and $\mathbold{d}_P^{*}$ are the optimal designs selected from the loss functions $\lambda_E(\mathbold{d,y})$ and $\lambda_P(\mathbold{d,y})$, respectively.

Figure \ref{fig:Est_E1} shows the parameter estimation performance of designs selected under each loss function. As expected, the estimation designs have the lowest value of expected loss within each scenario.  Of note is the fact that the dual-purpose designs are generally highly efficient when $n=10$, and this efficiency appears to increase with the spatial dependency. The prediction only designs (selected from $\tilde{\lambda}_P{(\mathbold{d,y})}$) did not perform well for the goal of parameter estimation, and this appears to become worse as the strength of spatial dependence increases. 

Figure \ref{fig:log_det_E1} shows the distribution of the expected loss values for different designs when the goal was prediction. As can be seen, the prediction only designs have the lowest expected loss when compared to the other designs. The  dual-purpose designs are highly efficient with respect to the prediction (only) designs.  Of note, this high efficiency is obtained despite the estimation designs on occasion being relatively inefficient.

Table \ref{tab:Eff_ex1} compares the efficiency of designs found under each experimental goal.  These results reflect those seen in the above figures in that the dual-purpose designs are efficient under each objective, and there are occasions when the estimation designs perform relatively poorly for prediction.  In contrast, the prediction designs maintain reasonable efficiency for estimation but not to the same extent as the dual-purpose designs.

 \vspace{-.4cm}
 \vspace{-.4cm}

 \vspace{-.4cm}

 \vspace{-0.3cm}

\begin{table}[H]
\begin{center} 
\caption{Efficiencies of designs selected under each loss function from Example 1} 
\begin{tabular}{lllrr}
\hline
\begin{tabular}[c]{@{}l@{}}Number of\\   design points\end{tabular} & \begin{tabular}[c]{@{}l@{}}Spatial\\   correlation\end{tabular} & \begin{tabular}[c]{@{}l@{}}Loss\\   function \end{tabular} & \begin{tabular}[c]{@{}l@{}}Estimation\\   efficiency (\%)\end{tabular} & \begin{tabular}[c]{@{}l@{}}Prediction\\  efficiency (\%)\end{tabular} \\ \hline
\multirow{9}{*}{5} & \multirow{3}{*}{Weak} & Estimation & 100.00 & 39.16 \\ 
 &  & Dual-purpose & 78.33 & 81.15 \\
 &  & Prediction & 70.76 & 100.00 \\ \cline{2-5} 
 & \multirow{3}{*}{Moderate} & Estimation & 100.00 & 66.75 \\ 
 &  & Dual-purpose & 77.92 & 81.27 \\
 &  & Prediction & 72.96 & 100.00 \\ \cline{2-5} 
 & \multirow{3}{*}{Strong} & Estimation & 100.00 & 49.62 \\  
 &  & Dual-purpose & 91.62 & 93.47 \\
 &  & Prediction & 76.58 & 100.00 \\ \cline{1-5} 
 \multirow{9}{*}{10} & \multirow{3}{*}{Weak} & Estimation & 100.00 & 77.72\\ 
 &  & Dual-purpose & 94.27 & 97.42 \\
 &  & Prediction & 70.21 & 100.00 \\ \cline{2-5} 
 & \multirow{3}{*}{Moderate} & Estimation & 100.00 & 78.58 \\ 
 &  & Dual-purpose & 93.43 & 98.01 \\
 &  & Prediction & 78.88 & 100.00 \\ \cline{2-5} 
 & \multirow{3}{*}{Strong} & Estimation & 100.00 & 75.61 \\  
 &  & Dual-purpose & 90.33 & 92.95 \\
 &  & Prediction & 74.65 & 100.00 \\ \cline{1-5} 
\end{tabular}
\label{tab:Eff_ex1}
\end{center}
\end{table} \vspace{-1.0 cm}
\subsection{Example 2: Motivating example revisited}

Here, we assess the performance of Algorithm 1 and the dual-purpose loss function in finding a retrospective design for Queensland's air quality monitoring network based on two pollutants $NO_{2}$ and $PM_{2.5}$. Similar to Example 1, $Y_1$ ($NO_{2}$) was assumed to follow a Normal distribution while $Y_2$ ($PM_{2.5}$) was assumed to follow a Poisson distribution as follows: \vspace{.5cm}
\[
Y_{1}|s_1\sim N(\mu_1,\sigma^2) \hspace{0.5 cm} \text{and} \hspace{0.5 cm} \mu_1= \beta_{10}+\beta_{11}X_1+\beta_{12}X_2+s_1,
\] 
\[
Y_{2}|s_2\sim Pois(\exp(\mu_2)) \hspace{0.5 cm} \text{and} \hspace{0.5 cm} \mu_2= \beta_{20}+\beta_{21}X_1+\beta_{22}X_3+s_2,
\]

where $X_1,X_2$ and $X_3$ are the Y-coordinate (in UTM system), annual mean humidity level, and the annual mean wind speed measured at the given location, respectively. Again, the Mat\'{e}rn covariance functions with parameters $(\mathbold{\gamma}_1, \nu_1)$ and $(\mathbold{\gamma}_2, \nu_2)$ were used to form each variance-covariance matrix to capture the spatial variability in each response.

To find the prior distributions for the parameters, we considered the data collected from the sampled locations over the period from 2013 to 2016. First, for each parameter, vague prior distributions were assumed. Then, the observed data were used to update this prior information to obtain a posterior distribution. It is this posterior distribution that was used for design selection as detailed in Algorithm 1. Here, the purpose of this experiment is to select an optimal set of monitoring stations to collect data so that they can be used to predict the annual $NO_2$ and $PM_{2.5}$ concentrations for all the monitoring stations in the network. As such, the optimal design is selected from all available locations across the network (i.e.\ both sampled and unsampled locations).

\begin{figure}[H]
	\centering
\includegraphics[width=11.0cm]{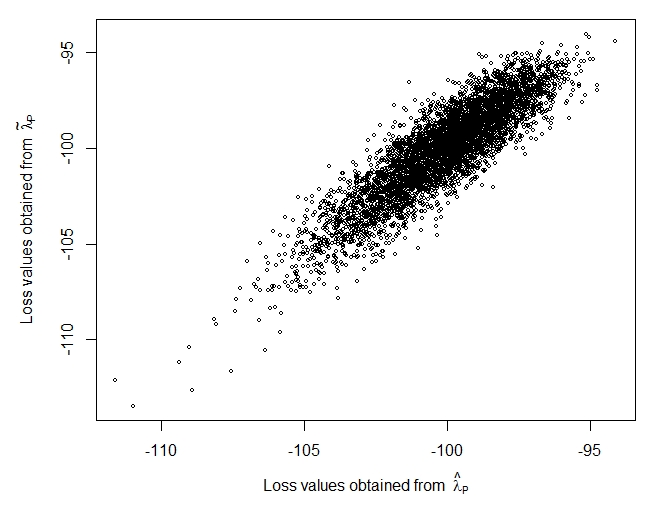} \vspace{-0.4cm}
	\caption{The relationship between the loss values obtained from $\hat{\lambda}_P(\mathbold{d,y})$ and $\tilde{\lambda}_P(\mathbold{d,y})$ for 100 randomly generated designs from Example 2}
	\label{fig:App_Ute_Ex2}
\end{figure}

\begin{figure}[H]
	\centering
\includegraphics[width=11.2cm]{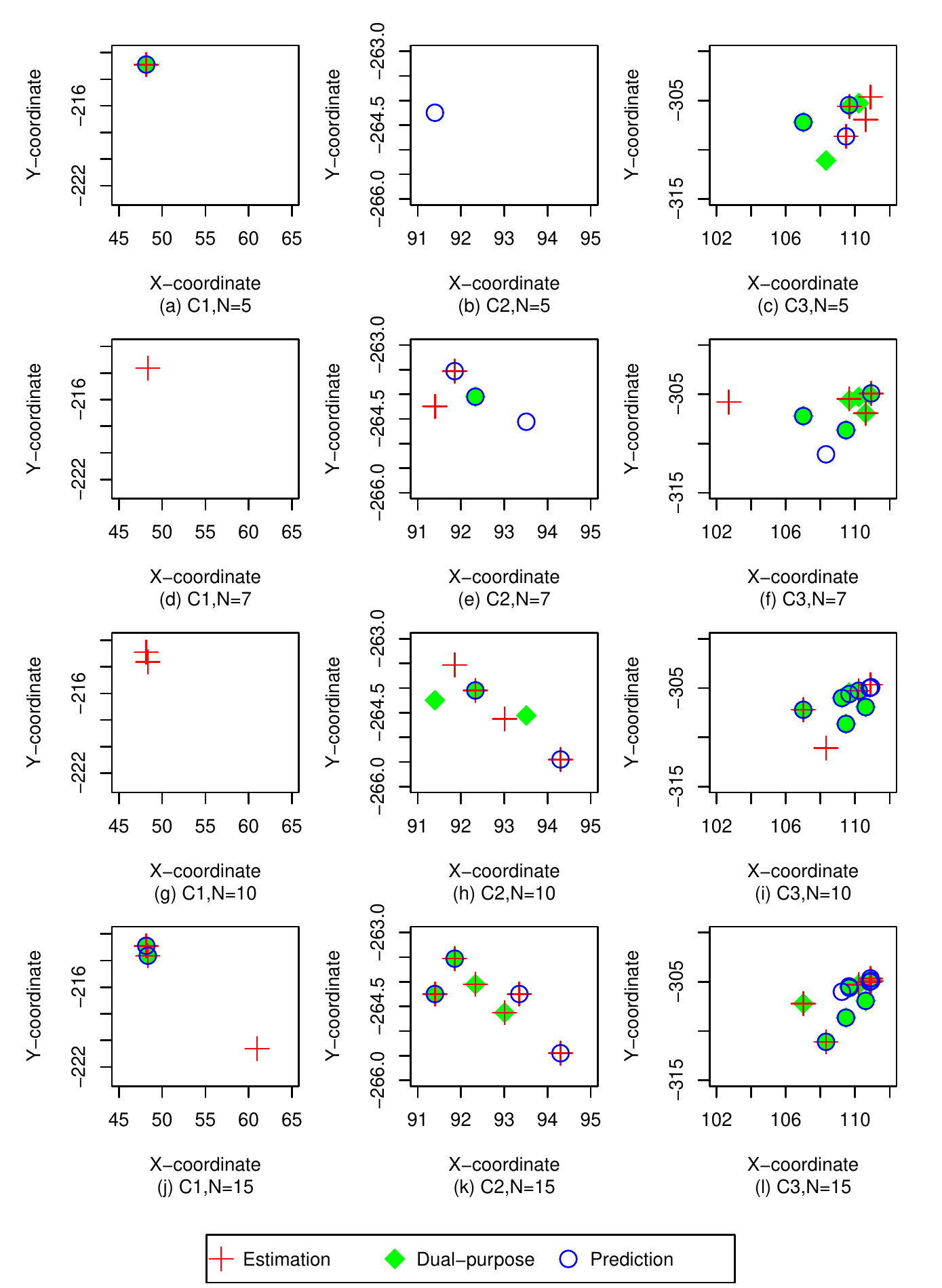} \vspace{-0.3cm}
	\caption{The optimal monitoring stations selected from each loss function}
	\label{fig:designs_Ex2}
\end{figure}

\textbf{Results:} Similar to the first example, we compared values of loss functions $\tilde{\lambda}_P(\mathbold{d,y})$ and $\hat{\lambda}_P(\mathbold{d,y})$ for 100 randomly generated designs each with 10 design points. Again, similar loss function values were obtained from each approximation, see Figure \ref{fig:App_Ute_Ex2}. As such, adopting our approximation to form efficient designs seems reasonable in this example. 
 \vspace{-.3 cm}
In Table \ref{tab:designs_Ex2}, we have summarised the designs found for different values of $n$ under each loss function. The numbers shown in this table correspond to the monitoring stations in Figure \ref{fig:site.loc}, with the bold numbers representing the sampled locations. Figure \ref{fig:designs_Ex2} shows the locations of the selected monitoring stations under each loss function. It can be seen that the designs selected under each loss function are different but have common stations. Further, when comparing the designs with 5 or 7 design points, it can be seen that the designs points selected from $\lambda_E(\mathbold{d,y})$ were spread over the three clusters (C1, C2 and C3) while the majority of design points selected from $\lambda_P(\mathbold{d,y})$ and $\lambda_D(\mathbold{d,y})$ were clustered in C3. Since a large number of predicted locations belonged to C3, to minimise the prediction uncertainty, it seems reasonable to select more design points from C3 when $\lambda_P(\mathbold{d,y})$ and $\lambda_D(\mathbold{d,y})$ were responsible for design selection. For the designs with 10 or 15 designs points, the prediction only design points were also spread over the three clusters while the majority of dual-purpose design points were clustered around sampled locations in C2 and C3.

\begin{figure}[H]
	\centering
\includegraphics[width=12.5cm]{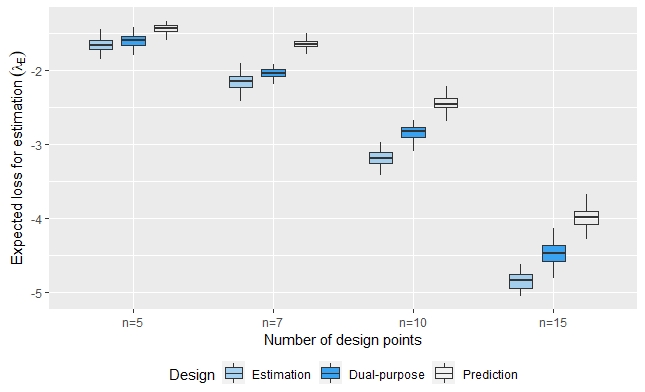} \vspace{-0.2cm}
	\caption{The boxplot of the distribution of the expected values for the loss function that focuses on parameter estimation for different designs over 50 simulations in Example 2}
	\label{fig:Est_Ex2}
\end{figure}

\begin{figure}[H]
	\centering
\includegraphics[width=12.5cm]{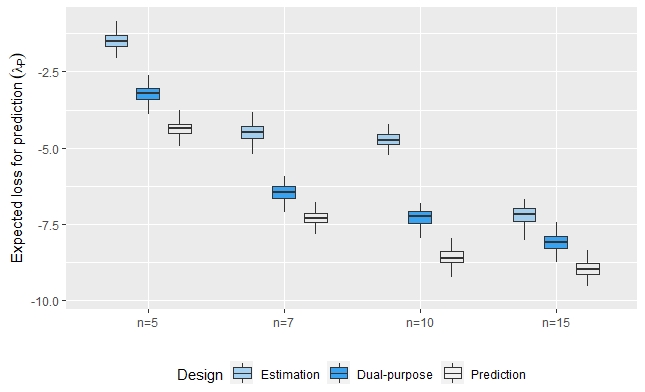} \vspace{-0.1cm}
	\caption{The boxplot of the distribution of the expected values for the loss function that focuses on prediction for different designs over 50 simulations in Example 2}
	\label{fig:log_det_Ex2}
\end{figure}

After locating the designs under each loss function, parameter estimation and prediction performance was assessed. Similar to Example 1, we considered 100 independent evaluations of the expected loss under each design objective. Figure \ref{fig:Est_Ex2} shows the distribution of values of the expected loss in terms of parameter estimation under different designs. Similar to Example 1, all estimation designs have lower expected loss when compared to other designs. As in the simulated example, the dual-purpose designs appear to be highly efficient in terms of parameter estimation, and this efficiency appears to increase with $n$.  The results also show that the relative performance of the prediction designs decreases as $n$ increases. Figure \ref{fig:log_det_Ex2} shows the distribution of expected loss values under the prediction loss function for different designs. As can be seen, the designs selected under the dual-purpose loss function perform well compared to the prediction only designs.  Further, the relative performance of the estimation designs for prediction appears to decrease with $n$. 

Table \ref{tab:Eff_ex2} shows the design efficiencies in terms of parameter estimation and prediction. Of note, the dual-purpose designs generally maintain high efficiency under both objectives.  For designs found under the other two objectives, there are occasions each perform relatively poorly, highlighted by efficiencies of less than 70\%.
\begin{table}[H]
\begin{center}
\caption {The optimal monitoring stations selected under each loss function} \label{tab:designs_Ex2} 
\begin{tabular}{clll} \hline
$n$ & Estimation design & Dual-purpose design & Prediction design \\ [0.6ex]  \hline \\[-1em]
5 & 15,21,27,31,33 & 8,11,14,31,33 & 2,11,21,{\bf{25}},31 \\ [0.9ex]
7 & 2,12,{\bf{25}},27,28,{\bf{29}},{\bf{34}} & 5,8,11,21,27,33,35 & 5,11,14,21,{\bf{26}},{\bf{29}},{\bf{34}} \\ [0.9ex]
10 & 5,{\bf{6}},8,{\bf{9}},11,14,15,28,{\bf{29}},31 & 2,5,8,11,21,{\bf{24}},{\bf{25}},{\bf{26}},27,33 & 5,{\bf{6}},8,11,21,{\bf{24}},27,33,{\bf{34}},35 \\ [0.9ex]
\multirow{2}{*}{15} & {1,2,3,5,\bf{6}},8,{\bf{9}},11,14,15, & 2,5,8,{\bf{9}},11,14,15,21,27,28, & 2,3,{\bf{6}},14,15,21,{\bf{24}},{\bf{25}},27,28, \\ & 28,{\bf{29}},31,{\bf{34}},35 & {\bf{29}},31,33,{\bf{34}},35 & {\bf{29}},31,33,{\bf{34}},35\\ [1.9ex]
\hline 
\end{tabular}
\end{center}
\end{table}

 \vspace{-.5cm}

 \vspace{-.5cm}

\begin{table}[H]
\caption{Efficiencies of designs selected under each loss function from Example 2} 
\begin{center}
\begin{tabular}{clrr}
\hline
\begin{tabular}[c]{@{}l@{}}Number of\\ design points\end{tabular} & \begin{tabular}[c]{@{}l@{}}Loss \\ function \end{tabular} & \begin{tabular}[c]{@{}l@{}}Estimation \\ efficiency (\%)\end{tabular} & \begin{tabular}[c]{@{}l@{}}Prediction\\ efficiency (\%)\end{tabular} \\ \hline
\multirow{3}{*}{5} & Estimation & 100.00 & 34.69 \\ 
 & Dual-purpose & 96.56 & 74.91 \\ 
 & Prediction & 85.11 & 100.00 \\ \hline
\multirow{3}{*}{7} & Estimation & 100.00 & 61.57 \\ 
 & Dual-purpose & 95.16 & 89.47 \\ 
 & Prediction & 75.99 & 100.00 \\ \hline
\multirow{3}{*}{10} & Estimation & 100.00 & 55.58 \\ 
 & Dual-purpose & 89.1 & 85.44 \\ 
 & Prediction & 76.01 & 100.00 \\ \hline
\multirow{3}{*}{15} & Estimation & 100.00 & 79.61 \\ 
 & Dual-purpose & 91.94 & 89.76 \\ 
 & Prediction & 82.46 & 100.00 \\ \hline
\end{tabular}
\label{tab:Eff_ex2}
\end{center}
\end{table} \vspace{-0.3 cm}
To further investigate future designs for monitoring air quality in Queensland, we re-ran this example by only considering the monitoring stations in C3. Optimal designs with 5 and 7 design points were selected using the three loss functions with the same prior distribution as detailed at the beginning of this example. The selected optimal designs are shown in Figure \ref{fig:designs_Ex2C3} in the Appendix A. Similar to the above results, the dual-purpose designs remain highly efficient for both design objectives (see Table \ref{tab:Eff_ex2C3}). Further, the relative performance of the single purpose designs appear to decrease as $n$ increases, see Figures \ref{fig:Est_Ex2C3} and \ref{fig:log_det_Ex2C3}.

\section{Discussion}

In this paper, we have developed a Bayesian design approach to minimise uncertainty in spatial predictions. The approach is based on the derivation of a dual-purpose loss function which quantifies the uncertainty in a given spatial process. Given the substantial computational time needed to evaluate this loss function, a number of approximations were proposed, and these were shown to work well for locating an optimal design. In addition, we extended our methodology to handle bivariate responses in spatial settings, and this was motivated by what has been observed in real-world studies. 

In the first example, three design scenarios were considered to test our proposed algorithm and the dual-purpose loss function under various levels of spatial dependence. Overall, the results showed that the dual-purpose designs remained highly efficient under each objective, and this was despite some single objective designs being inefficient under the other objective. Similar results were observed when designing Queensland's air quality monitoring network based on two air quality measurements $NO_2$ and $PM_{2.5}$. The clustered nature of the stations (in terms of spatial locations) was accounted for via the range parameter in our covariance function which resulted in the three clusters being independent. Alternative approaches and/or covariance functions to account for this clustering could be considered in the future, and this may have implications in terms of the design.

Despite the theoretical underpinnings of the loss functions developed in this study, the computational complexity of the loss function $\lambda_P(\mathbold{d,y})$ is a drawback of this design approach. Indeed, this difficulty led to the development of an efficient approximation based on the joint distribution of the parameters and the data. We evaluated the proposed approximation to the loss functions, in terms of the accuracy and the computational time required to obtain the approximate loss values. A comparison of the computational times required to evaluate each approximation is shown in Figures 14 and 15. This was based on evaluating 500 independent utilities on a computer with an Intel core-i7 processor running at  2.2 Gigahertz. For the models considered in this paper, this was shown to yield a reasonable approximation. However, such an approach may not be appropriate in general. For example, if bivariate continuous and binary data were observed, it is unlikely that a Multivariate Normal approximation will be appropriate for estimating the entropy of the distribution of the data. As such, alternative methods are needed, and this is an area we plan to explore into the future.

Another possible extension to this work could include quantifying the uncertainty in different components of the spatial process. This can be achieved by quantifying the uncertainty in the mean model and the covariance function, potentially following the work of \cite{borth1975total, mcgree2016developments}. The proposed loss function could then consider a set of all plausible mean and covariance functions, which is flexible enough to represent the uncertainty in the spatial process \citep{Pilz1997}. In studies where high-dimensional multivariate data are observed, it may be necessary to adopt particular
Vine-Copulas \citep{brechmann2013modeling, Panagiotelis2017138}. Other areas we hope to pursue into the future include extensions to quantify uncertainty in temporal (and thus, spatio-temporal) processes \citep{LIU2020100392}. 

\section*{Acknowledgement}

SGJS was supported by QUTPRA scholarship from the Queensland University of Technology. WGM was partially supported by Austrian Science Fund (FWF): I 3903-N32. JMM was supported by an Australian Research Council Discovery Project (DP200101263). Computational resources and services used in this work were provided by the HPC and Research Support Group, Queensland University of Technology, Brisbane, Australia.

\bibliographystyle{agsm}
\small {
}


\section*{Appendix A} 
\begin{figure}[H]
	\centering
\includegraphics[width=11cm]{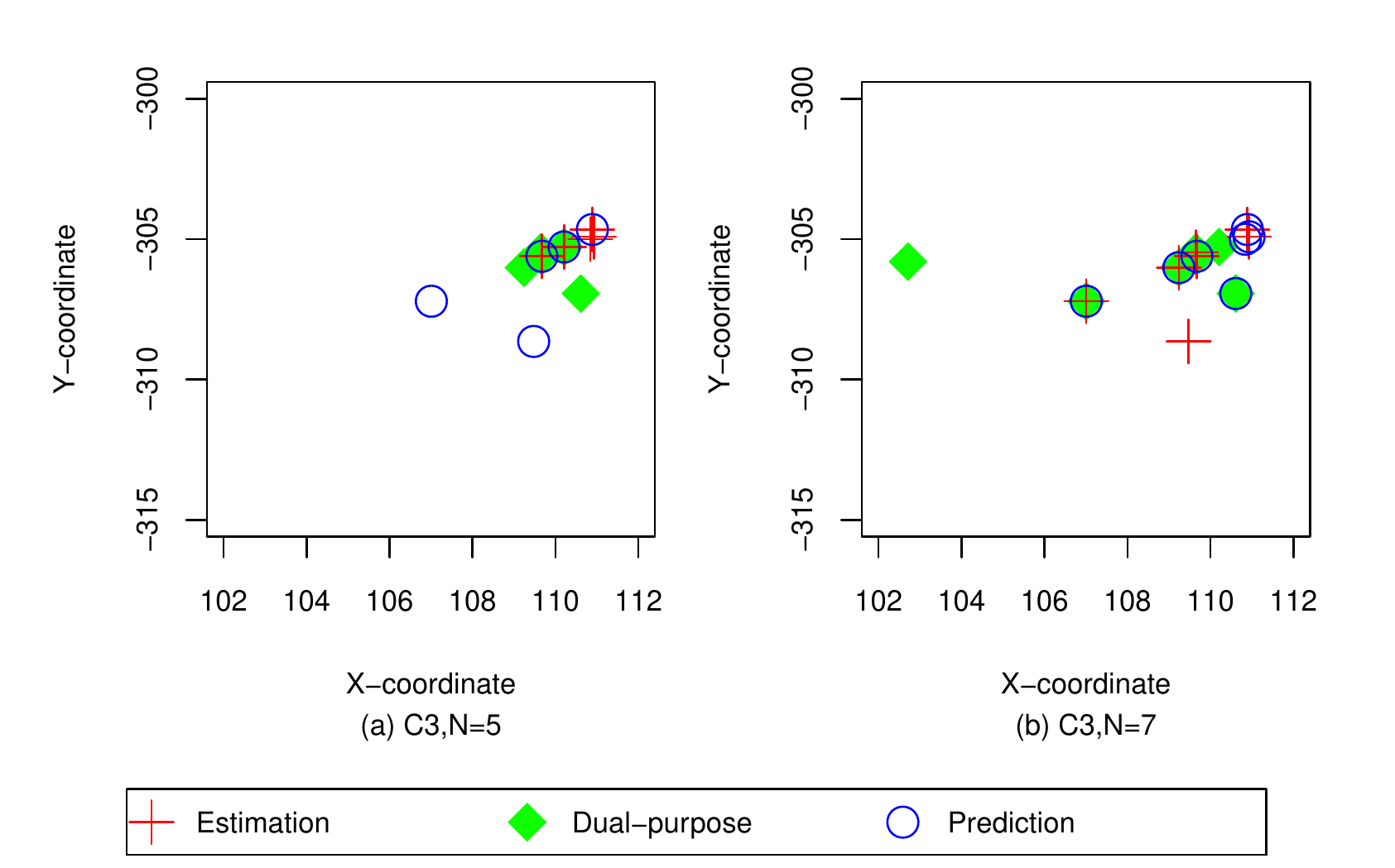} \vspace{-0.1cm}
	\caption{The optimal monitoring stations selected from C3 in Example 2 using each loss function}
	\label{fig:designs_Ex2C3}
\end{figure} \vspace{-0.3cm}

\begin{figure}[H]
	\centering
\includegraphics[width=11cm]{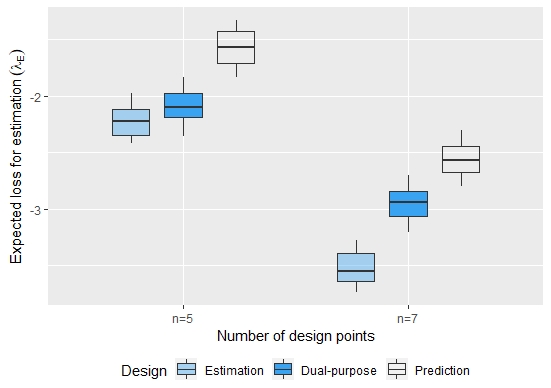} \vspace{-0.1 cm}
	\caption{The boxplot of the distribution of the expected values for the loss function that focuses on parameter estimation for different designs over 50 simulations from C3 in Example 2}
	\label{fig:Est_Ex2C3}
\end{figure} 

\begin{figure}[H]
	\centering
\includegraphics[width=11cm]{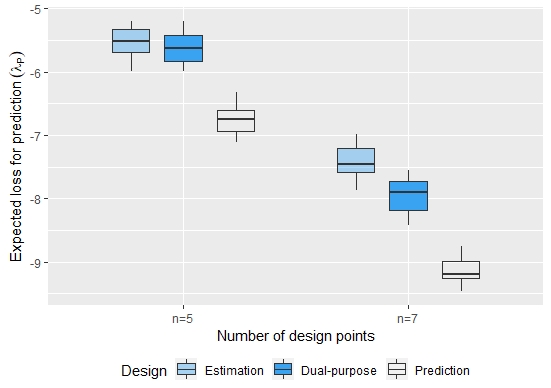} \vspace{-0.1 cm}
	\caption{The boxplot of the distribution of the expected values for the loss function that focuses on prediction for different designs over 50 simulations from C3 in Example 2}
	\label{fig:log_det_Ex2C3} 
\end{figure} 

\begin{table}[H]
\caption{Efficiencies of designs selected under each loss function from C3 in Example 2} 
\begin{center}
\begin{tabular}{clrr}
\hline
\begin{tabular}[c]{@{}l@{}}Number of\\ design points\end{tabular} & \begin{tabular}[c]{@{}l@{}}Loss \\ function \end{tabular} & \begin{tabular}[c]{@{}l@{}}Estimation \\ efficiency (\%)\end{tabular} & \begin{tabular}[c]{@{}l@{}}Prediction\\ efficiency (\%)\end{tabular} \\ \hline
\multirow{3}{*}{5} & Estimation & 100.00 & 80.62 \\ 
 & Dual-purpose & 93.56 & 82.88 \\ 
 & Prediction & 71.73 & 100.00 \\ \hline
\multirow{3}{*}{7} & Estimation & 100.00 & 82.03 \\ 
 & Dual-purpose & 83.13 & 87.34 \\ 
 & Prediction & 72.13 & 100.00 \\ \hline
\end{tabular}
\label{tab:Eff_ex2C3}
\end{center}
\end{table} \vspace{-.1 cm}

\begin{figure}[H]
	\centering
\includegraphics[width=12cm]{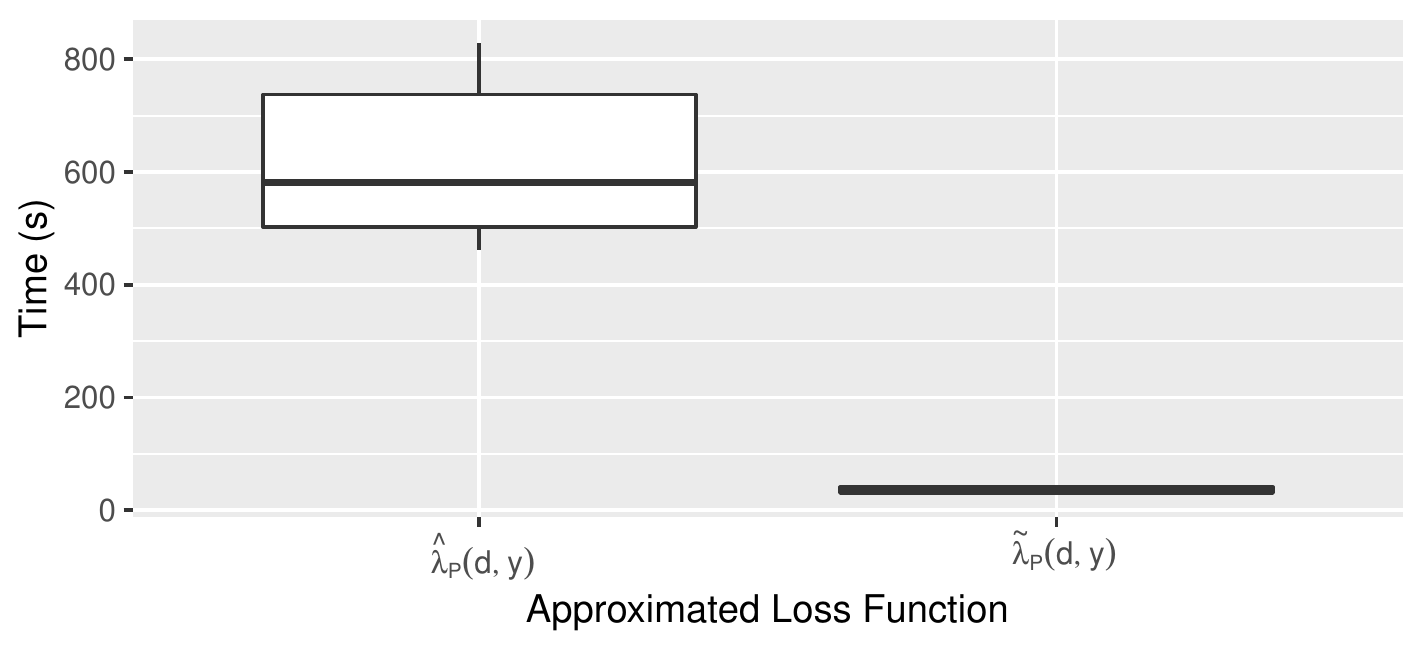} \vspace{-0.1cm}
	\caption{The comparison of computational time required to estimate the two loss functions for Example 1}
	\label{fig:Time_Ex1}
\end{figure} \vspace{-.3 cm}

\begin{figure}[H]
	\centering
\includegraphics[width=12cm]{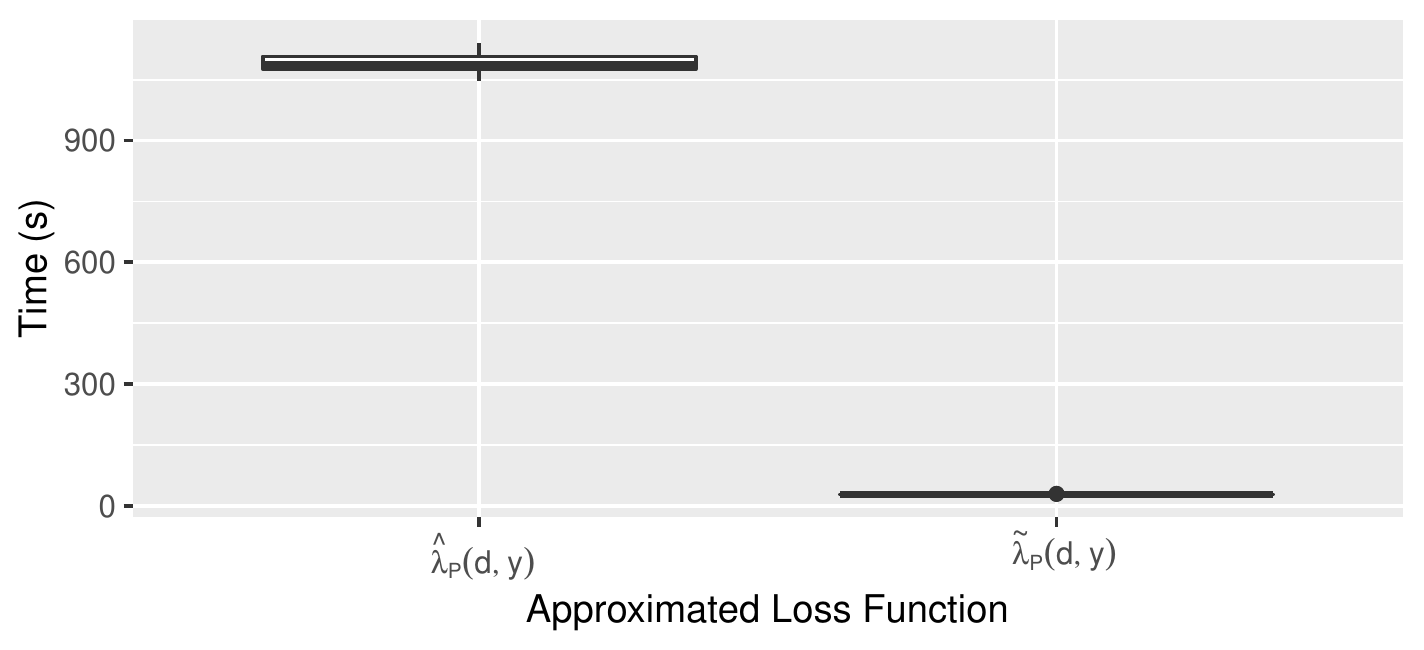} \vspace{-0.1cm}
	\caption{The comparison of computational time required to estimate the two loss functions for Example 2}
	\label{fig:Time_Ex2}
\end{figure} 

\section*{Appendix B}

In spatial settings, the literature suggests that parameter estimation designs often feature clustered sampling points \citep{Digglebook2007}. Here the ratio of the range and nugget to the partial sill influence the configuration of clusters. In contrast, the prediction designs often resemble a regularly spaced design. It is presumed that designs for both of these goals will combine features from both.

\begin{figure}[H]
	\centering
\includegraphics[width=11.3cm]{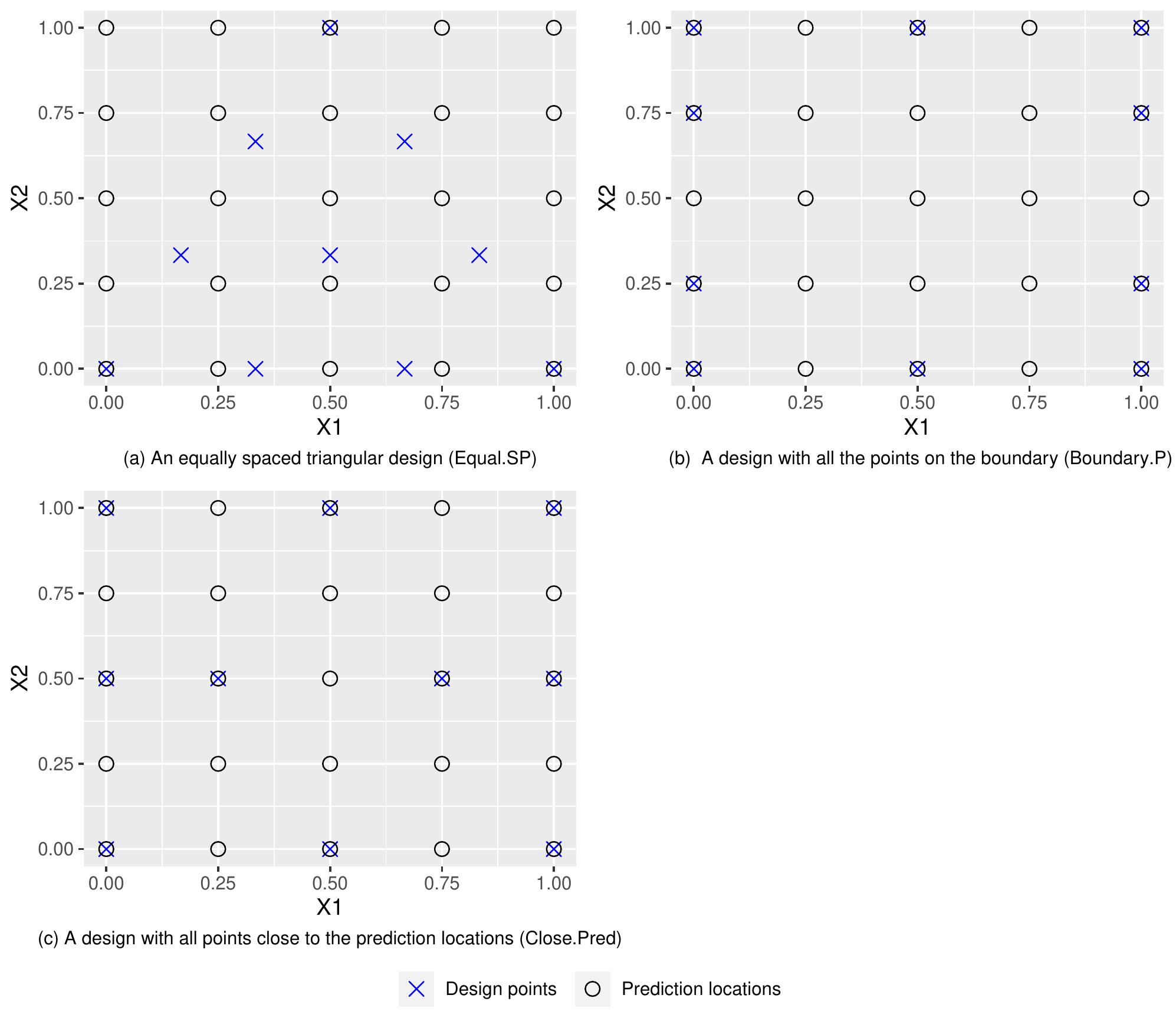} 
	\caption{A selected set of theoretical designs}
	\label{fig:Theoretical_d}
\end{figure}

In Example 1, we potentially observe some of the above features across the three different designs with some apparent difference. To investigate about these differences, a new simulation study has been conducted where the estimation and prediction performance of a selected set of designs were evaluated. These designs were: an equally spaced triangular design, a design with all the points on the boundary, and a design with all points close to the prediction locations, see Figure \ref{fig:Theoretical_d} below which shows these three designs. The performance of these designs for estimation and prediction were compared to the designs found in this work. 

  \vspace{-.3cm} 

\begin{figure}[H]
	\centering
\includegraphics[width=11.0cm]{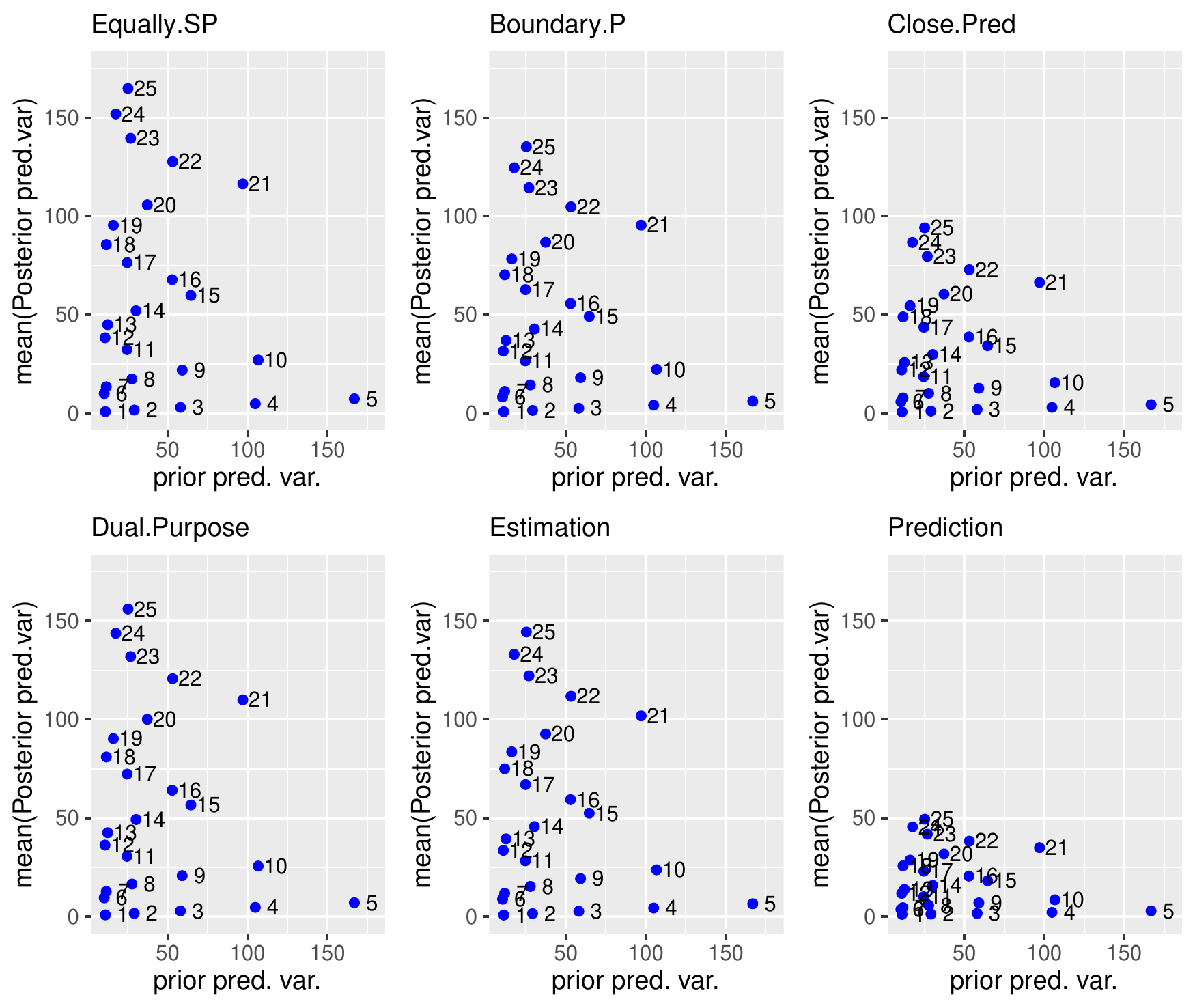} \vspace{-0.1cm}
	\caption{Posterior prediction variances vs prior prediction variances of response 1 (Y1) for prediction locations}
	\label{fig:Post_pred_var_Y1}
\end{figure}

\begin{figure}[H]
	\centering
\includegraphics[width=11.0cm]{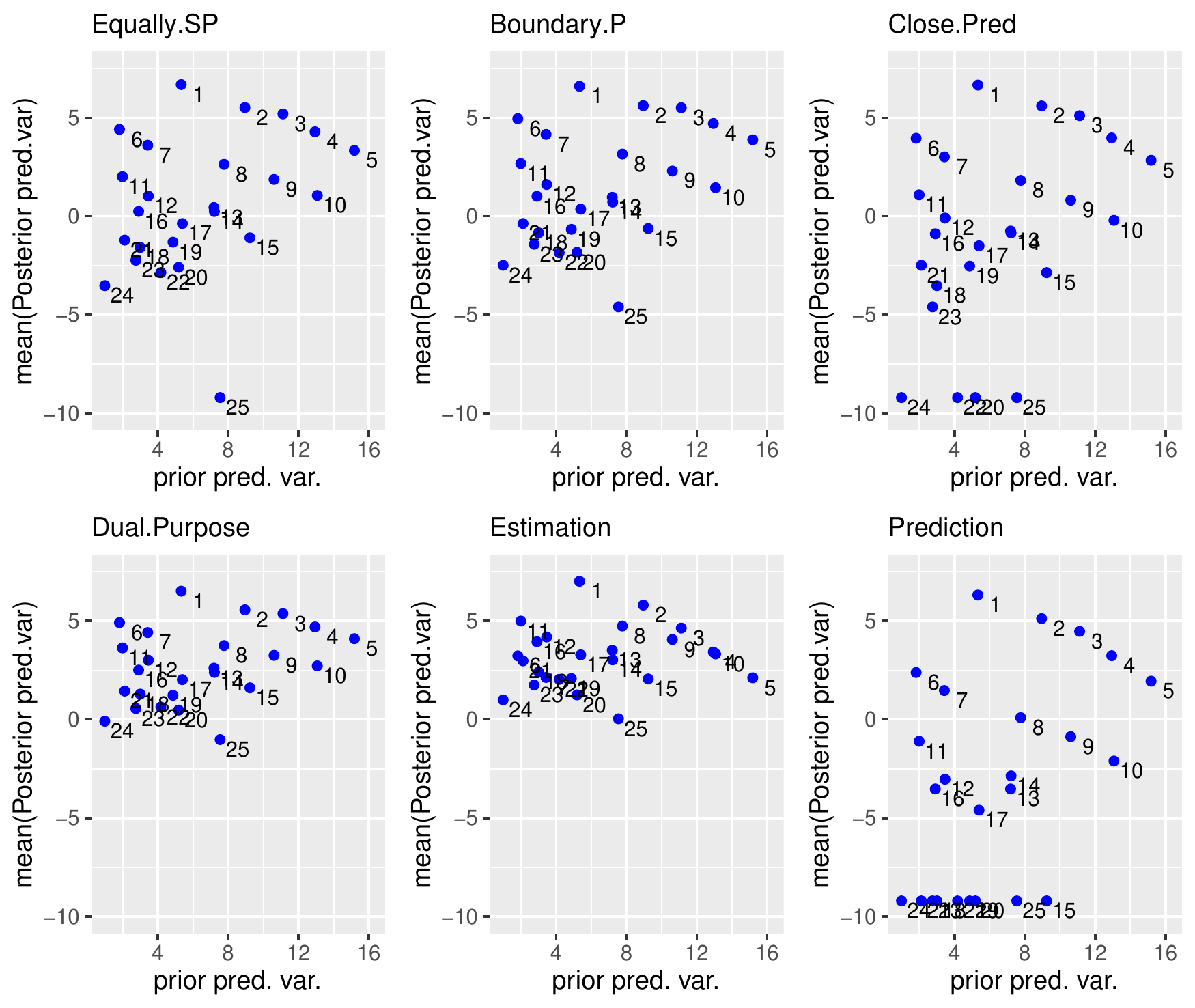} \vspace{-0.1cm}
	\caption{Posterior prediction variances vs prior prediction variances of response 2 (Y2) for prediction locations}
	\label{fig:Post_pred_var_Y2}
\end{figure}

To evaluate the prediction performance of the designs shown in Figure \ref{fig:Theoretical_d}, we evaluated the expected prior and posterior prediction variance at each of the prediction locations.  This was achieved by simulating 500 data sets from each design and approximating the posterior distribution for each. For each of these posterior distributions, posterior predictive data were simulated at each prediction location, and then the variance of these at each location was evaluated.  The average of this variance was then evaluated across the 500 simulations.  The results for this for both responses are shown in Figures \ref{fig:Post_pred_var_Y1} and \ref{fig:Post_pred_var_Y2}.

From Figure \ref{fig:Post_pred_var_Y1}, it is clear that the prediction design found in our paper yields the smallest average prediction variance across all locations. The next best design appears to be the ‘close.pred’ design suggesting that a design that is more regularly spaced across the entire region performs well here for prediction, and this is what would be expected based on results from the literature. Otherwise, there appears to be very little difference in the performance of the remaining designs. Given one of these designs in the estimation design found in our work, then potentially these designs are focussing more on estimation.

From Figure \ref{fig:Post_pred_var_Y2}, similar conclusions can be drawn.

\vspace{-0.5cm}

\begin{figure}[H]
	\centering
\includegraphics[width=12.5cm]{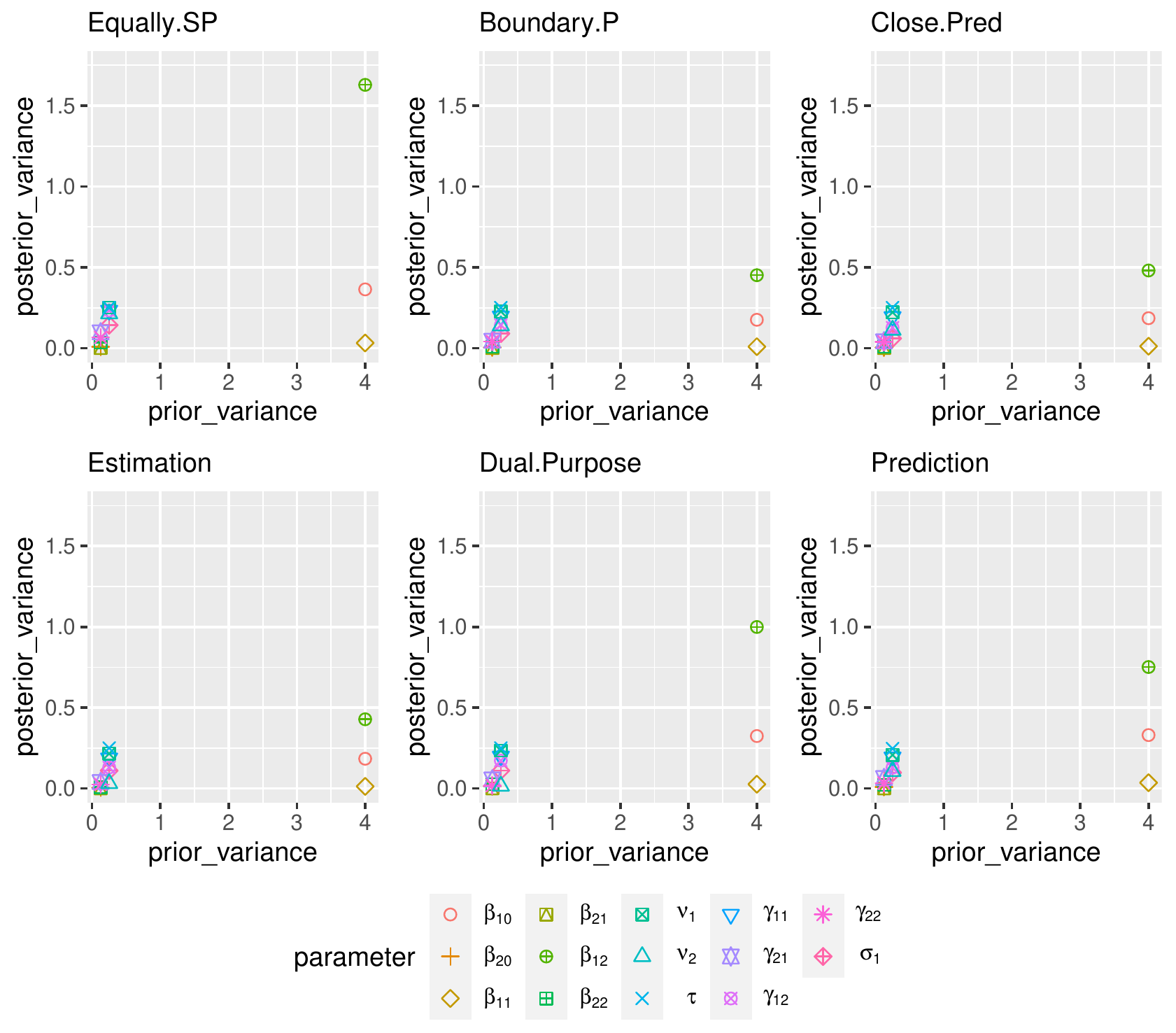} 
	\caption{The prior variance vs the posterior variance for each parameter for different designs}
	\label{fig:Post_Para_variance}
\end{figure}

To evaluate the estimation performance of the designs, a similar approach to evaluate the prediction performance was followed. That is, 500 prior predictive data sets were generated based on each design, and a posterior distribution was found for each. The average posterior variance of each parameter was then evaluated based on each design. These averages are shown in Figure \ref{fig:Post_Para_variance}. Across all designs, there does not appear to be an appreciable difference in terms of estimation performance. However, our estimation design is performing better but only marginally. This may be because the prior variance for most parameters is quite low, so attention may be given to estimating $\beta_{11}, \beta_{12}$ and $\beta_{10}$ more precisely. To do so, at least for the slope terms, a contrast between data collected at extreme values of X1 and also X2 would seem useful, and this is a feature of all designs considered.

The above highlights the potentially influence the prior distribution can have in Bayesian design, and provide a potential reason for differences observed between our designs and those that might be considered as being more standard for estimation and prediction in spatial settings. Given this, such standard designs may not necessarily be optimal under a variety of prior configurations. This was observed here but we also note that these standard designs did not perform poorly.

\end{document}